\documentclass[aps,prb,twocolumn,groupaddress,showpacs,floatfix]{revtex4}

\usepackage[bookmarks]{hyperref}
\usepackage[dvips]{graphicx}
\usepackage{amsmath}
\usepackage{amstext}
\usepackage{graphics}
\usepackage{exscale}
\usepackage{epsfig}
\usepackage[T1]{fontenc}
\usepackage{bm}
\usepackage{bbm}
\usepackage{color}

\usepackage{version}

\usepackage{latexsym}

\usepackage[tight]{subfigure}

\renewcommand{\epsilon}{\varepsilon}

\newcommand{\integral}[3]{\!\int\limits_{#2}^{#3}\!\!{\rm d}#1\;}

\newcommand{\expval}[2]{ \langle  #1 #2\ \!\! \rangle}

\newcommand{\e}{\mathrm e}

\newcommand{\vct}[1]{\bm #1}
\newcommand{\vk}{{\bm k}}
\newcommand{\vq}{{\bm q}}

\newcommand{\Imag}{\mathrm{Im}}
\newcommand{\Real}{\mathrm{Re}}

\includeversion{commentst1} 
\begin{document}

\title{Real space Eliashberg approach to charge order of
  electrons\\ coupled to dynamic antiferromagnetic fluctuations}  
\author{Johannes Bauer${}^{1}$ and Subir Sachdev${}^{1,2}$}
\affiliation{${}^{1}$Department of Physics, Harvard University, Cambridge,
  Massachusetts 02138, USA}
\affiliation{${}^{2}$Perimeter Institute for Theoretical Physics, Waterloo, Ontario N2L 2Y5, Canada}
%\author{Subir Sachdev}
%\email[]{jbauer@physics.harvard.edu}
%\affiliation{Department of Physics, Harvard University, Cambridge,
%  Massachusetts 02138, USA}
\date{\today} 

\begin{abstract}
We study charge ordered solutions for fermions on a square lattice
interacting with dynamic antiferromagnetic fluctuations. Our approach is
based on real space Eliashberg equations which are solved
self-consistently. We first show that the antiferromagnetic fluctuations can 
induce arc features in the spectral functions, as spectral weight is
suppressed at the hot spots; however, no real pseudogap is generated. At low temperature spontaneous charge
order with a $d$-form factor can be stabilized for certain
parameters. As long as the interacting Fermi surfaces possesses hot spots, the
ordering wave vector corresponds to the diagonal connection of the hot spots,
similar to the non-self-consistent case. Tendencies towards observed axial order only appear
in situations without hot spots.  
\end{abstract}

\maketitle

\section{Introduction} 

Signatures of charge order in the copper-oxide based superconductors have
attracted a lot of recent attraction. By now charge ordered states have become
an essential ingredient of the phase diagram for the different families.\cite{FK12,KKNUZ15} 
%(such as Bi, Y, Hg-based) 
Whilst in La-based compounds charge and spin stripe order have a considerable
history,\cite{TSANU95,EKT99,KBFOTKH03} it was only firmly established in
recent years in other cuprate  families. Early reports include scanning
tunneling microscopy (STM) studies,\cite{HHLMEUD02,VMOAAY04,Kea07,Lea10} followed
by bulk property measurements by resonant X-ray scattering
(REXS).\cite{Gea12,Cea12,Aea12} The order has a finite correlation length in zero 
magnetic field and becomes long range ordered at high magnetic fields.\cite{Wea11,LKHKBP13} This helps to
understand previously puzzling quantum oscillation data. \cite{Dea07,SHL12,ACS14}

By now the charge order has been carefully characterized experimentally. It is
unconventional in the sense that it possess an incommensurate wave length
$\lambda=3-4a$, where $a$ is the Cu-Cu distance in the copper oxide planes and
an internal form factor. In both STM and REXS the direction of the wave
vector has been established as pointing along the axes in the Cu-Cu square
lattice, $\vct Q=Q_0(1,0)$.\cite{Gea12,Cea12,Cea14a,Dea14a} Its magnitude
follows the Fermi surface, or more precisely, Fermi arc 
evolution; concretely this means that it decreases with increasing hole
doping.\cite{Cea14a} This suggests that the order is connected to Fermi surface
properties and can possibly be understood as a Fermi surface
instability.\cite{Cea14a,Dea14a}  In contrast, in La-based compounds the
ordering wave vector shows the opposite trend, and $\vct Q$ increases with
increasing hole doping.\cite{Hea11,Tra13} Recent experiments have revealed further
details of the charge order in Bi- and Y-based compounds. Both STM and REXS
data are well understood based on a $d$-form factor as an internal structure
for the order.\cite{Fea14,Cea15,Hea15pre}

\begin{figure}
\includegraphics[width=180pt]{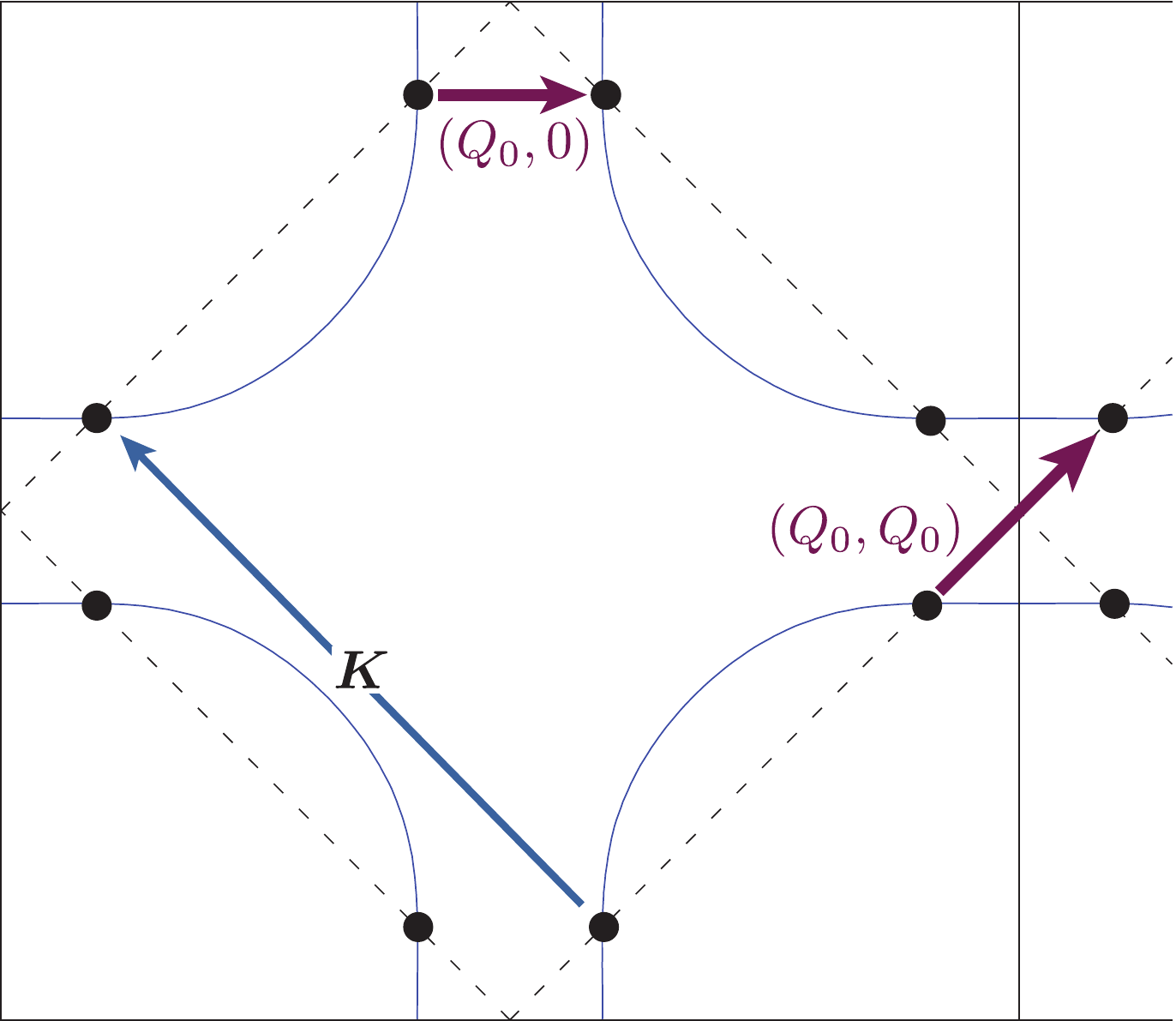}
\caption{(Color online) Schematic picture of Fermi surface with
  antiferromagnetic wave vector connecting hot spots $\vct K$ and examples of
  diagonal, $\vct Q=Q_0(1,1)$, and axial, $\vct Q=Q_0(1,0)$, charge ordering wave vectors.}
\label{fig:1}
\vspace*{-0.5cm}
\end{figure}

%$t_1 = 1$, $t_2 = -0.32$, $t_3=0.128$, and $\mu=-1.119$. For this dispersion we have $Q_0
%  \simeq 0.36$ in units of $\pi/a$

Many theories have addressed the microscopic origin of the charge order
including its onset temperature, its wave vector, form factor and dependence on doping and magnetic
field.\cite{HM12a,HM12b,BGY12,SL13,KKNUZ15,WC14,MS14,TC14,Lee14,DF14,MS14,MN15pre,WC15} One line of reasoning starts with fermions interacting with
antiferromagnetic spin fluctuations peaked at wave vector $\vct K$ (see
Fig.~\ref{fig:1}).\cite{MS10a,SL13,SS14,ABS14a,ABS14b}
The hot spot model with linearized dispersion has a strong nesting instability
with a diagonal wave vector $\vct Q=Q_0(1,1)$ connecting the hot 
spots (see Fig.~\ref{fig:1}).\cite{MS10a} In models starting with an intact Fermi surface
this instability is at least as strong as the one with an axial wave vector $\vct Q=Q_0(1,0)$, and it
is therefore difficult to explain the experiments based on such a theory. 
A number of recent works have explored situations where the non-interacting
Fermi surface has been modified due to magnetic order and pocket 
formation\cite{BAK13,AKB15,TS15} or pseudogap features.\cite{ZM14pre}
It has been argued that it is important to take such features into account,
such that the dominant instability can be altered. 
Also fluctuation effects,\cite{CS14a,PCKM14,WAC15,Pun15} strong correlation
effects,\cite{ABS14a,ABS14b} and starting points of fractional Fermi
liquids\cite{CS14b} have been considered as a possibility to favor axial order
over the diagonal one.  

The main idea of this work is to analyze whether single particle self-energy
modification, due to a coupling to dynamic spin fluctuations, and without appeal
to more exotic states, is sufficient to establish charge order with properties as observed in
experiments. Our analysis will be carried out with a semi-phenomenological model of fermions
coupled to dynamic spin fluctuations, also called spin-fermion model in the
literature. Such a model is appealing as it naturally captures the $d$-wave
superconductivity in a range of dopings.\cite{ACS03,CPS08} 
We show that in the relevant doping regime for charge order, dynamic magnetic fluctuations
with a finite correlation length $\xi\sim 2a$ are a more
realistic description for the cuprates than including static magnetic
order.\cite{BAK13,AKB15}  
To analyze the model, we use a real space formulation which allows us to probe
charge order directly in the ordered phase. Our formulation also allows for
simple extensions to include impurities or disorder, spin order, coupling to
phonons, superconducting order, and a magnetic field. 
The paper is structured as follows: In Sec.~II we describe the details of the
model and approach. In Sec.~III we analyze ordered solutions in the static
limit in order to connect to previous work. In Sec.~IV we analyze the full
dynamic model before concluding in Sec.~V.

\section{Model definition and real space Eliashberg approach}
\label{sec:mod}
We use a model with the action of the form,\cite{ACS03,CPS08}
\begin{eqnarray}
  S&=&-\sum_{\sigma}\sum_{ij,n} \overline{\psi}_{i,\sigma}(i\omega_n)
  G_{ij,0}^{-1}(i\omega_n)\psi_{j,\sigma}(i\omega_n) \label{eq:spfmod}\\
&-&\frac{g^2}{2}\sum_{i,j}\integral{\tau}{}{}\!\!\!\integral{\tau'}{}{}\chi(\vct r_i-\vct r_j,\tau-\tau')
  \vct S(\vct r_i,\tau)\cdot \vct S(\vct r_j,\tau'), \nonumber
\end{eqnarray}
where
\begin{equation}
  G_{ij,0}(i\omega_m)^{-1}=(i\omega_m+\mu)\delta_{ij}-t_{ij},  
\end{equation}
and
\begin{equation}
  S^{\alpha}(\vct r_i,\tau)=\overline{\psi}_{i,\sigma_1}(\tau)\sigma^{\alpha}_{\sigma_1,\sigma_2}\psi_{i,\sigma_2}(\tau).
\end{equation}
Here, $\psi_{i,\sigma}$ is a fermionic field for site $i$ and spin label $\sigma$.
The non-interacting dispersion in momentum space is
\begin{eqnarray*}
  \epsilon_{\vk}&=&-2t_1[\cos(k_1)+\cos(k_2)]-4t_2\cos(k_1)\cos(k_2) \\
&& -2t_3[\cos(2 k_1)+\cos(2 k_2)],
\label{eq:dispersion}
\end{eqnarray*}
where the copper-copper lattice spacing is set to $a=1$. 

For the spin-fluctuation spectrum we assume,\cite{MMP90,ACS03,CPS08}
\begin{equation}
  \chi(\vq,i\omega_m)=\frac{a_{\chi}}{N_{K}}\sum_{i=1}^{N_K}\frac{1}{\omega_{\vct  K_i}(\vq)+a_{v_s}^2
    \omega_m^2 +\frac{|\omega_m|}{\omega_{\rm sf}}}.
\label{eq:chidyn}
\end{equation}
To preserve lattice periodicity we define $\omega_{\vct K_i}(\vq)=2 (2-\cos(q_x-\vct K_{x,i})-\cos(q_y-\vct
K_{y,i}))+\Gamma^2$, and we usually take for simplicity $\vct K=(\pi,\pi)$.
$\Gamma$ is directly related to the correlation length $\Gamma=\xi^{-1}$. 
For $a_{v_s}=0$ this form is common in the literature such as in the
well-known work of Millis {\em et al.\/}\cite{MMP90} and fits experimental data from
neutron scattering well.
In real space we have 
\begin{equation}
  \chi(\vct r_i-\vct r_j,i\omega_m)=\frac{1}{N_s}\sum_{\vq} \chi(\vq,i\omega_m)
  \e^{i \vq (\vct r_i-\vct r_j)},
\label{eq:chi}
\end{equation}
where $N_s$ is the number of lattice sites.

\paragraph{Real space equations -}
The basic equation for the Eliashberg approach is
\begin{equation}
  \Sigma_{ij}(i\omega_n)= 3 T g^2 \sum_m\chi(\vct
  r_i-\vct r_j,i\omega_n-i\omega_m) G_{ij}(i\omega_m),
\label{eq:re}
\end{equation}
where $G$ is the full Green's function. We assume that no spin order occurs
and omit spin labels on $G$ and $\Sigma$. In order to compute $\Sigma$
self-consistently we also need to solve the Dyson equation,
\begin{equation}
  G_{ij}(i\omega_m)^{-1}=(i\omega_m+\mu)\delta_{ij}-t_{ij}-h_{ij}-\Sigma_{ij}(i\omega_m).
\label{eq:dysonre}
\end{equation}
When working in real space $G_{ij}(i\omega_m)$ can be obtained by matrix
inversion for each $i\omega_m$, a computation which can be easily
parallelized. The calculation of $\Sigma$ in Eq.~(\ref{eq:re}) can also
be parallelized well.
In the Green's function $G$ in Eq.~(\ref{eq:dysonre}), we have included a
general symmetry breaking field, which can be introduced 
into the Hamiltonian by 
\begin{equation}
  H_{\rm co}=\sum_{i,j,\sigma}h_{ij}c_{i,\sigma}^{\dagger}c_{j,\sigma}.
\label{eq:Hco}
\end{equation}
In order to find inhomogeneous solutions we have to initialize the
calculations with such a field.
We assume that $\chi$ in Eq.~(\ref{eq:chidyn}) is fixed by the input parameters and not further
renormalized in a self-consistent manner.
We work on a two-dimensional lattice with $N_s=N_1\times N_1$ sites and
periodic boundary conditions.

\paragraph{Homogeneous case -}
In the homogeneous situation $G$ and $\Sigma$ only depend on $\vct r_i-\vct
r_j$. With a Fourier transform we can then write,
\begin{equation}
  \Sigma_{\vk}(i\omega_n)=3g^2T\sum_{m,\vq}\chi(\vk-\vq,i\omega_n-i\omega_m) G(\vq,i\omega_m)
\label{eq:sigmaiw}
\end{equation}
and the Dyson equation can be solved explicitly,
\begin{equation}
  G_{\vk}(i\omega_m)=\frac{1}{i\omega_m-\epsilon_{\vk}+\mu-\Sigma_{\vk}(i\omega_m)}.
\end{equation}
%The spectral functions reads $\rho_G(\vk,\omega)=-\frac1{\pi}\Imag G_{\vk}(\omega+i\eta)$.
In the spectral representation we have
\begin{eqnarray*}
  \Sigma_{\vk}(i\omega_n)&=&3g^2\sum_{\vq}\integral{\omega_1}{}{}\!\!\integral{\omega_2}{}{}\frac{\rho_{\chi}(\vk-\vq,\omega_1)
    \rho_G(\vq,\omega_2)}{i\omega_n-\omega_1-\omega_2}  \\
&&   \times [n_{\rm  F}(\omega_2)+n_{\rm B}(-\omega_1)] ,
\end{eqnarray*}
where $\rho_{\chi}(\vq,\omega)$ is the spectral function for the
spin-fluctuation spectrum, and $n_{\rm  F}(\omega)$, $n_{\rm B}(\omega)$ denote
the fermionic and bosonic distribution functions, respectively. Analytic
continuation, $i\omega\to \omega+i\eta$, yields the imaginary 
part of the self-energy 
\begin{eqnarray}
 \Imag \Sigma_{\vk}(\omega)&=&-3\pi g^2\sum_{\vq}\integral{\omega_2}{}{}
 \rho_{\chi}(\vk-\vq,\omega-\omega_2)   \rho_G(\vq,\omega_2)  \nonumber \\
&& \times [n_{\rm   F}(\omega_2)+n_{\rm B}(\omega_2-\omega)] .
\label{eq:imsigmaw}
\end{eqnarray}
and the real part can be computed from the Kramers-Kronig transformation. From
this the electronic spectral function $\rho_G(\vk,\omega)$ can be computed. 
For a numerical calculation it is favorable to use the real space form
for the self-energy equation and switch to momentum space by Fast Fourier
transforms.

\begin{figure*}[!ht]
%\vspace*{-0.5cm}
\centering
\includegraphics[width=0.48\linewidth]{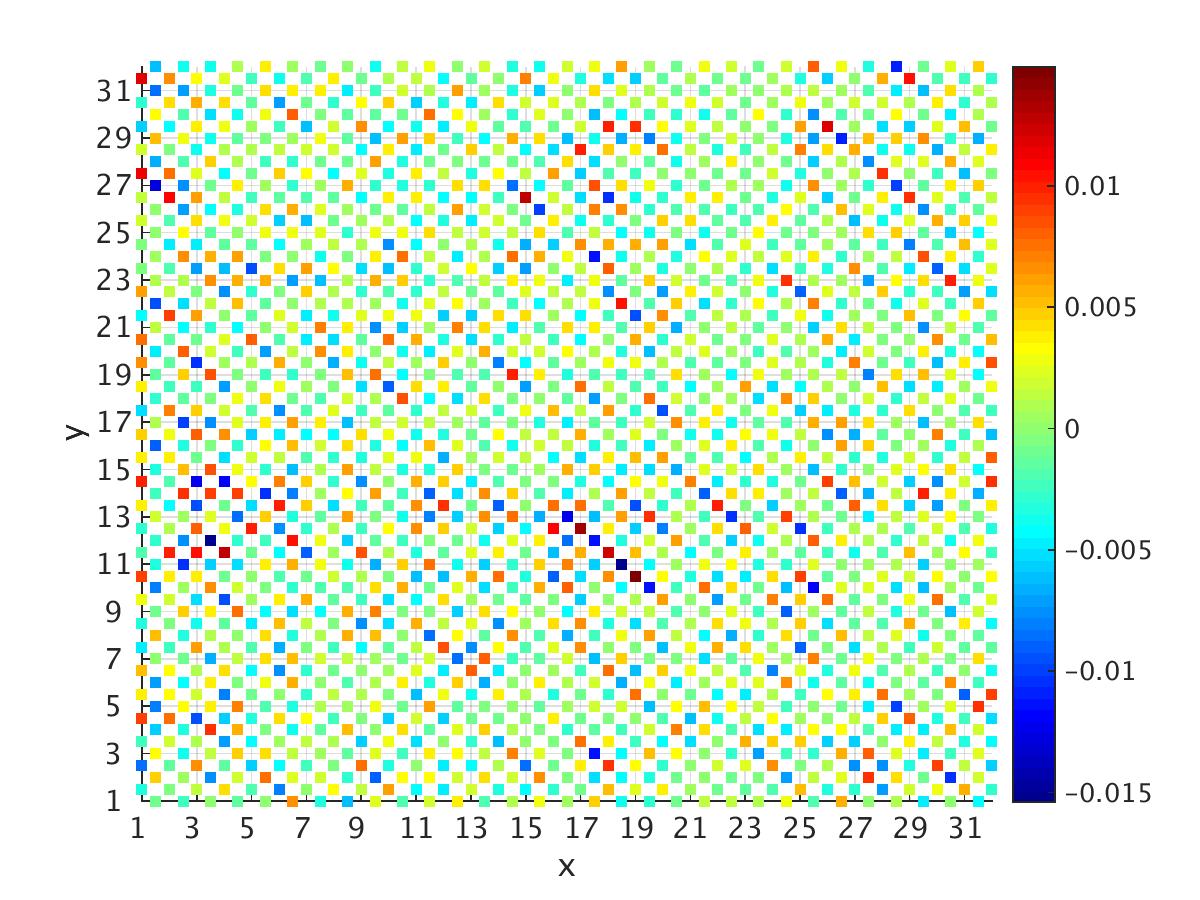}
\includegraphics[width=0.48\linewidth]{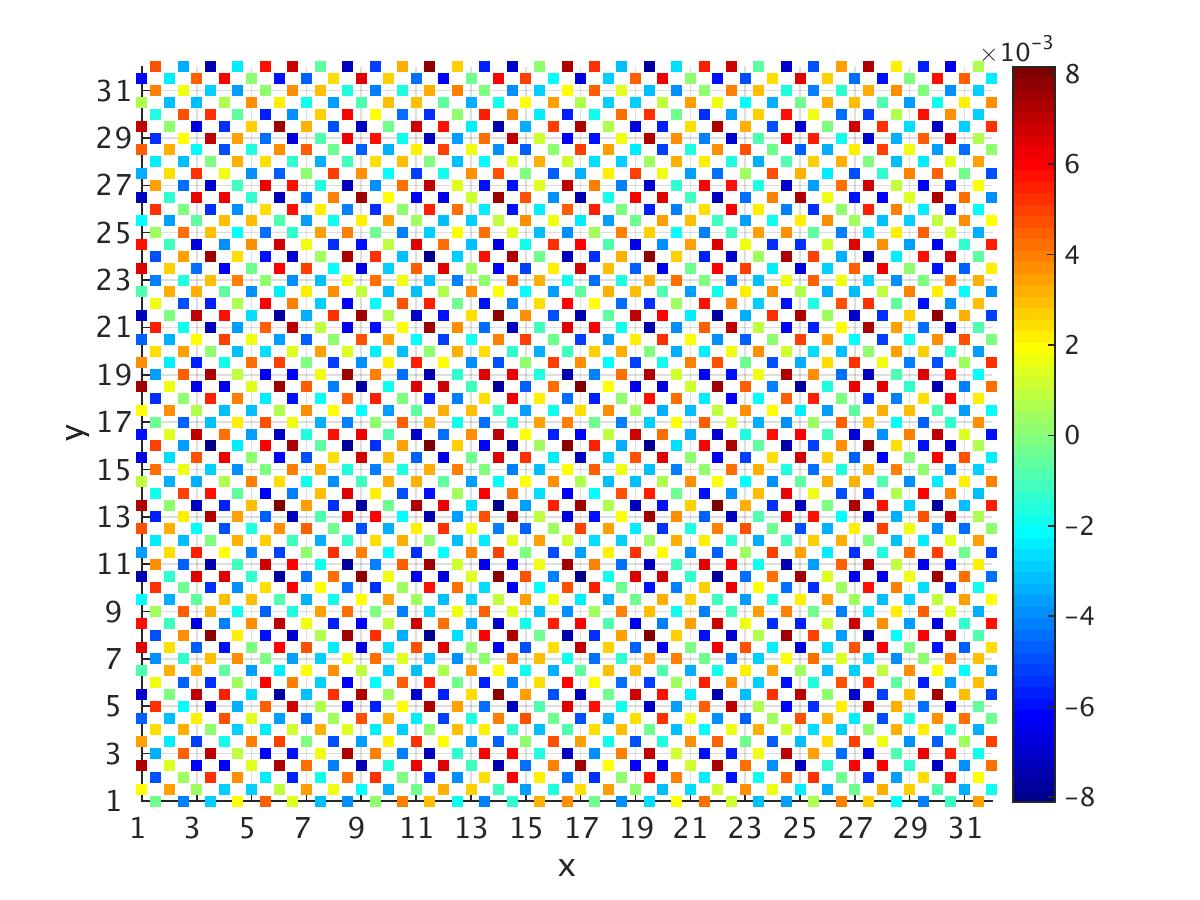}
\vspace*{-0.2cm}
%\vspace*{-0.3cm}
\caption{(Color online) Results for the static model. We plot the nearest
  neighbor bond values of $\Delta_{ij}^{\rm  ih}$ on a real space lattice
  initialized by a random field after 5  iterations (left) and spontaneously
  ordered results after 30 iterations (right).}            
\label{fig:2}
\vspace*{-0.2cm}
\end{figure*}

\paragraph{Charge order observables -}
For the order parameter we focus on the inhomogeneous part of 
\begin{equation}
  \Delta_{ij}=\sum_{\sigma}\expval{c_{i,\sigma}^{\dagger}c_{j,\sigma}}{}=\sum_{\vct
    Q}\Big[\frac{1}{V}\sum_{\vk}\Delta_{\vct Q}(\vk)\e^{i\vk \vct r}\Big]\e^{i\vct Q \vct R}.
\end{equation}
where $\vct r=\vct r_i-\vct r_j$ and $\vct R=(\vct r_i+\vct r_j)/2$.
Given a spin-diagonal Green's function $G_{ij}(i\omega_n)$ we can compute
$\Delta_{ij}=2T\sum_n G_{ij}(i\omega_n)\e^{i\omega_n\tau^-}$, $\tau^-\to 0$. To extract the
inhomogeneous part $\Delta^{\rm ih}$ we map it to coordinates $\vct R$, $\vct r$,
$\Delta_{ij}\to \Delta(\vct R,\vct r)$ and then compute $\Delta(\vct
r)=\Delta(\vct R,\vct r)-\frac{1}{N_R}\sum_{\vct  R}\Delta(\vct R,\vct
r)$. Then we find $\Delta^{\rm ih}(\vct R,\vct r)=\Delta(\vct R,\vct
r)-\Delta(\vct r)$. Without any charge order symmetry breaking $\Delta^{\rm ih}$ is zero.

We assume that the inhomogeneous part $\Delta^{\rm ih}_{\vct  Q}(\vk)$ can be
expanded as
\begin{equation}
  \Delta^{\rm ih}_{\vct  Q}(\vk)=\sum_n a_n(\vct Q) f_n(\vk),
\end{equation}
with a suitable set of orthonormal basis functions $\{f_n(\vk)\}$. The
explicit form of the relevant functions used here can be found in the appendix. 
These can be transformed to the real space representation,
\begin{equation}
  f_n(\vct r)=\frac{1}{N_s}\sum_{\vk}f_n(\vk)\e^{-i\vk \vct r}.
\label{eq:fnr}
\end{equation}
For a given function $\Delta^{\rm ih}(\vct R,\vct r)$, the coefficients
$a_n(\vct Q)$ can be calculated from 
\begin{equation}
  a_n(\vct Q)= \sum_{\vct R,\vct r}\Delta^{\rm ih}(\vct R,\vct r) \e^{-i\vct Q
    \vct R} f_n(\vct r).
\label{eq:anQ}
\end{equation}

\paragraph{Parameters of the model -}
The model in Eq.~(\ref{eq:spfmod}) contains a number of parameters, which we summarize here for
clarity. The bare electronic structure is determined by the hoppings $t_1$,
$t_2$ and $t_3$, and the filling $n=1/N_s\sum_{i,\sigma}n_{i,\sigma}$ by the
chemical potential $\mu$. Unless otherwise mentioned $t_1=1$ sets the energy
scale. To get a rough idea about absolute scales we can think of a typical
identification $t_1\simeq 300$meV$\simeq 3481$K; however, we make no attempt
for a quantitative theory in comparison with experiment here.
The spin fluctuation spectrum has the following parameters: the
overall weight factor $a_{\chi}$, the inverse correlation length
$\Gamma=\xi^{-1}$, the spin fluctuation scale $\omega_{\rm sf}$ and the 
$\omega^2$ coefficient $a_{v_s}$.  We use parameters similar to the recent work of Mishra and
Norman.\cite{MN15pre} In addition we have the temperature $T$, where the 
lowest value reached is  $T=0.02 t_1$ ($\simeq 70$K). Moreover we have a coupling constant
$g$. A technical parameter is the size of the real space lattice. We did most
of our calculations for $N_1=32$. The limit for this is set by memory
constraints.

\section{Results for the static model}
\label{sec:static}
We first consider the situation where the spin-fluctuation spectrum is a
static, which corresponds to the limit $\omega_{\rm sf}\to \infty$ 
and $a_{v_s}\to 0$ in Eq.~(\ref{eq:chidyn}). This helps us to connect to previous
results\cite{SL13} and test our formalism and procedure. 
The equation for the self-energy then simplifies to,
\begin{equation}
\Sigma_{ij}=3 g^2 \chi(\vct r_i-\vct r_j)\frac{\Delta_{ij}}{2}.
\end{equation}
This purely static renormalization does not lead to any quasiparticle
damping. However, it leads to a sizeable renormalization of the chemical
potential and hopping parameters. Using the relative coordinate $\vct r=\vct
r_i-\vct r_j$, we can relate the renormalized parameters $\{\mu,t_i \}$ to the
bare ones 
$\{\mu_0,t_i^0\}$ by $\mu=\mu_0+\Sigma(\vct r=0)$, $t_1=t_1^0+\Sigma(|\vct r|=1)$
$t_2=t_2^0+\Sigma(|\vct r|=\sqrt{2})$, and $t_3=t_3^0+\Sigma(|\vct r|=2)$,
where $t_i^0$ are the bare parameters.

In our procedure we search for spontaneously ordered solutions of the self-consistency equations 
(\ref{eq:re}) and (\ref{eq:dysonre}) by initializing the calculations  by a
random field $h_{ij}$ [see  
Eq.~(\ref{eq:Hco})]. This is initially included in the Green's function in
Eq.~(\ref{eq:dysonre}) and then set to zero from the second iteration on. We
use a real space lattice with $N_1=32$ sites in one direction and iteratively
calculate the full Green's function $G_{ij}$ and self-energy
$\Sigma_{ij}$. Some mixing of iterations and initial 
onsite homogenization is used to improve convergence. 
One example for a charge order solution is obtained with the parameter set
$t_1^0=0.764$, $t_2^0=-0.33$, $t_3^0=0.154$, and $\mu_0=0.393$. For
$T=0.05$, $\Gamma=\xi^{-1}=0.5$, and $g^2=3$, we obtain $t_1\approx 1.0$, $t_2\approx
-0.32$, $t_3\approx 0.11$, $\mu\approx-1.1$, and the renormalized Fermi
surface looks very similar to Fig.~\ref{fig:1}.  From the converged result for $G_{ij}$,
the expectation values $\Delta_{ij}$ and $\Delta^{\rm ih}_{ij}$ are obtained
as described in Sec.~\ref{sec:mod}. In Fig.~\ref{fig:2}, we plot the result for the nearest neighbor
bonds in $x-$ and $y-$direction as obtained from $\Delta^{\rm ih}_{ij}$. This
shows how an ordering pattern spontaneously appears and is stabilized after
30 iterations. 

\begin{figure}[htbp]
%\vspace*{-0.5cm}
\centering
\includegraphics[width=0.48\textwidth]{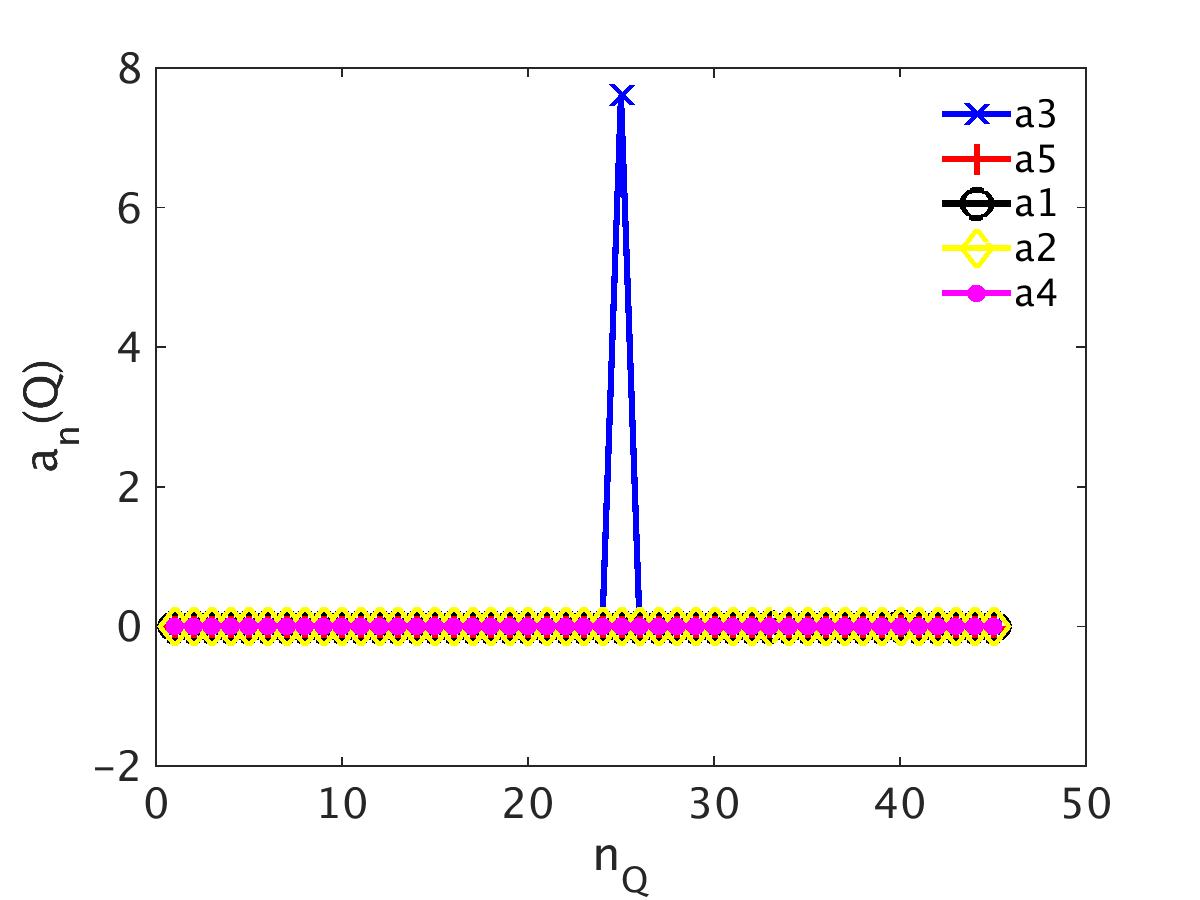}
\vspace*{-0.2cm}
\vspace*{-0.3cm}
\caption{(Color online) Plot of the largest $a_n(\vct Q)$ for different basis functions
  ($n=1,\ldots, 13$). The values of $\vct Q$ are labeled by $n_{\vct Q}$, where
  $n_{\vct Q}=25$ corresponds to $\vct Q=Q_0(1 ,1)$, with $Q_0=3/8$ in units
  $\pi/a$. This value agrees well with the diagonal distance between hot
  spots.}           
\label{fig:3}
\vspace*{-0.3cm}
\end{figure}

The properties of the ordering pattern can either be identified directly from the real space
representation in Fig.~\ref{fig:2} or well understood by the decomposition into the basis
function, Eq.~(\ref{eq:anQ}).  The largest coefficients are shown in 
Fig.~\ref{fig:3}. For simplicity, we only show the coefficients for a series
of momenta $\vct Q=(Q_x,Q_y)$ in the triangle, $Q_x\in [0,\pi]$,
$Q_y\in [0,Q_x]$, labeled by $n_Q$. 

There is a clear maximum for $a_3$ for $n_{\vct Q}=25$ which corresponds to
$\vct Q=Q_0(1 ,1)$, with $Q_0=3/8=0.375$ in units $\pi/a$. This value agrees
well with the diagonal distance between the hot 
spots on the Fermi surface, and the ordering form factor is of the d-wave
form, $n=3$, $f_3(\vk)=\cos k_x - \cos k_y$ (see
Table~\ref{tab:basisfunc} in the appendix). As mentioned we only show the result for a
restricted set of $\vct Q$ vectors. The order in Fig.~\ref{fig:1} is really a
superposition of $\pm (Q_x,Q_y)$, $\pm (Q_x,-Q_y)$ wave vectors.
We conclude that our real space Eliashberg calculations are consistent with
earlier work based on an instability analysis.\cite{SL13} The $\vk$-space
resolution for the finite size lattice is sufficient to resolve these
features and the relation to the Fermi surface. The dominant instability is
realized here as an ordered solution.

\begin{figure}[!ht]
%\vspace*{-0.5cm}
\centering
\includegraphics[width=0.48\linewidth]{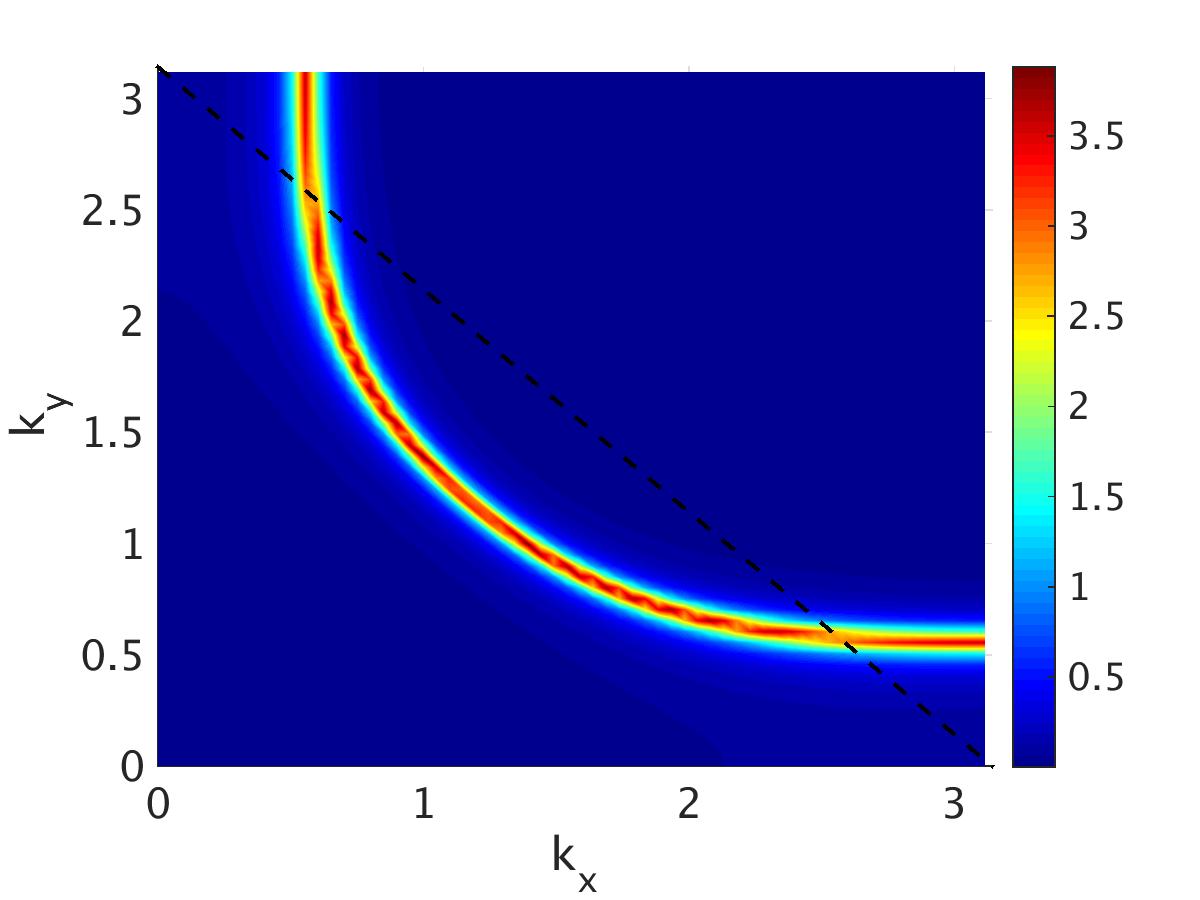}
\includegraphics[width=0.48\linewidth]{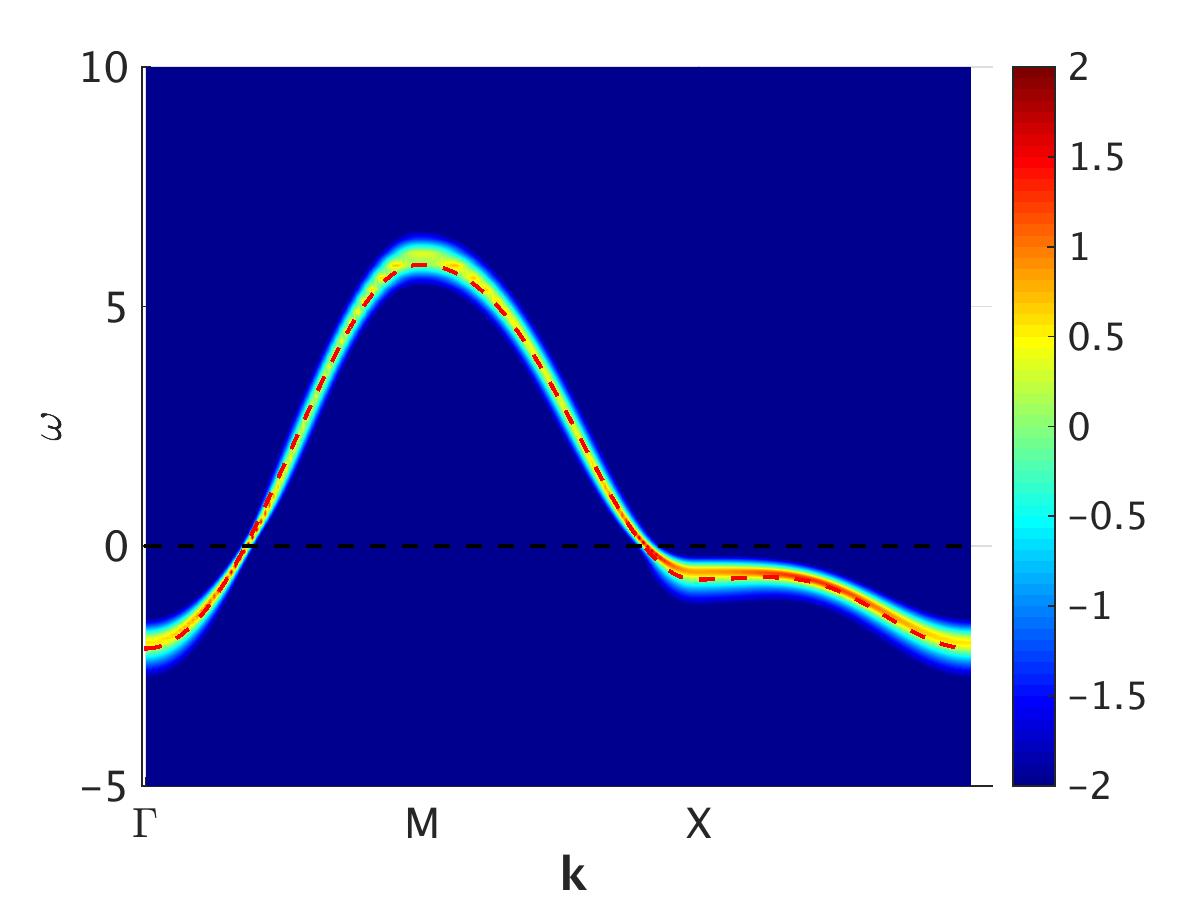}
\vspace*{-0.2cm}
%\vspace*{-0.3cm}
\caption{(Color online) Plot of interacting Fermi surface in the dynamic model with
weak coupling, $g=1.0$. We plot the spectral density
  ($\rho_{\vk}(\omega=0)$) (left) and
  renormalized band structure ($\log\rho_{\vk}(\omega)$ for clarity) for $\vk$
  along the trajectory $\Gamma\to M \to X \to \Gamma$ (right). The dashed line
  gives the non-interacting dispersion. The parameter values are given in the text.}            
\label{fig:4}
\vspace*{-0.2cm}
\end{figure}

\begin{figure*}[!ht]
%\vspace*{-0.5cm}
\centering
\includegraphics[width=0.32\linewidth]{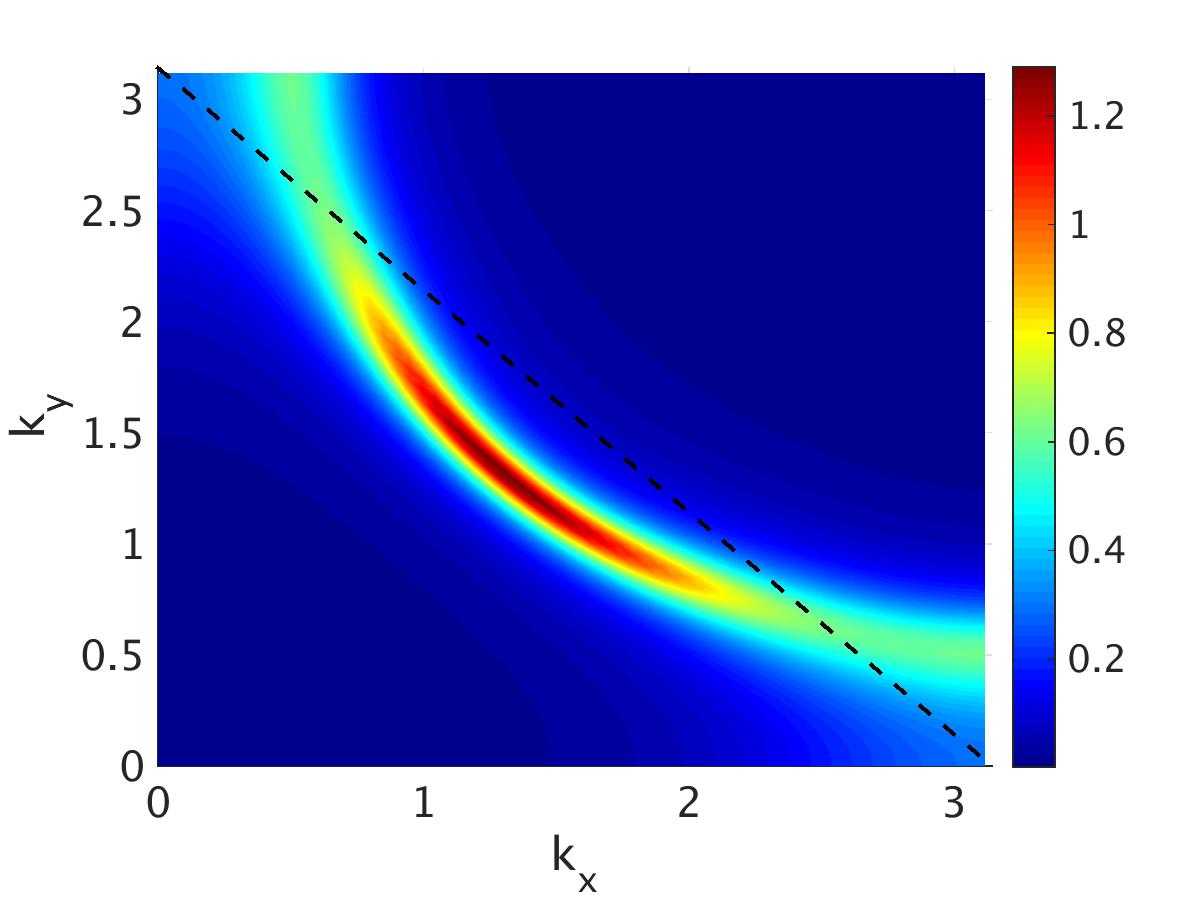}
\includegraphics[width=0.32\linewidth]{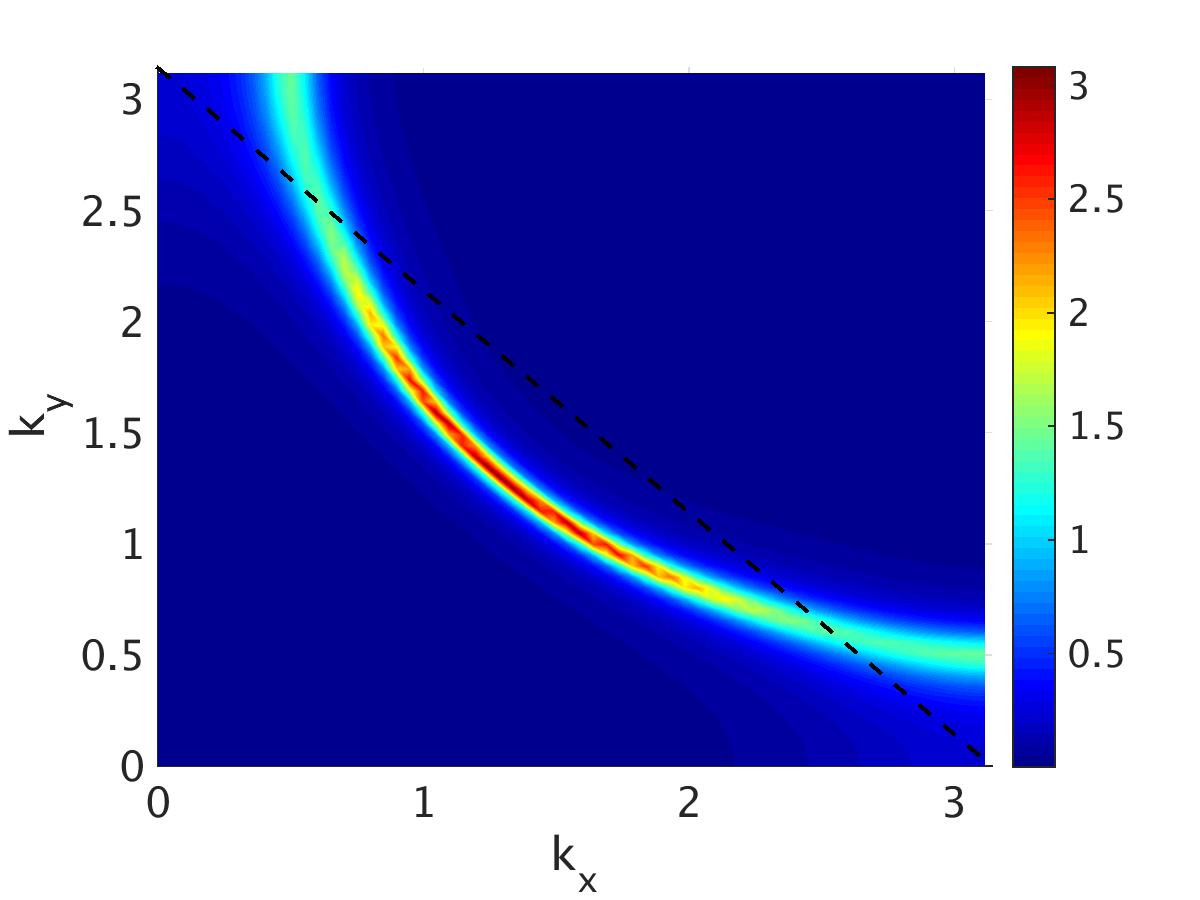}
\includegraphics[width=0.32\linewidth]{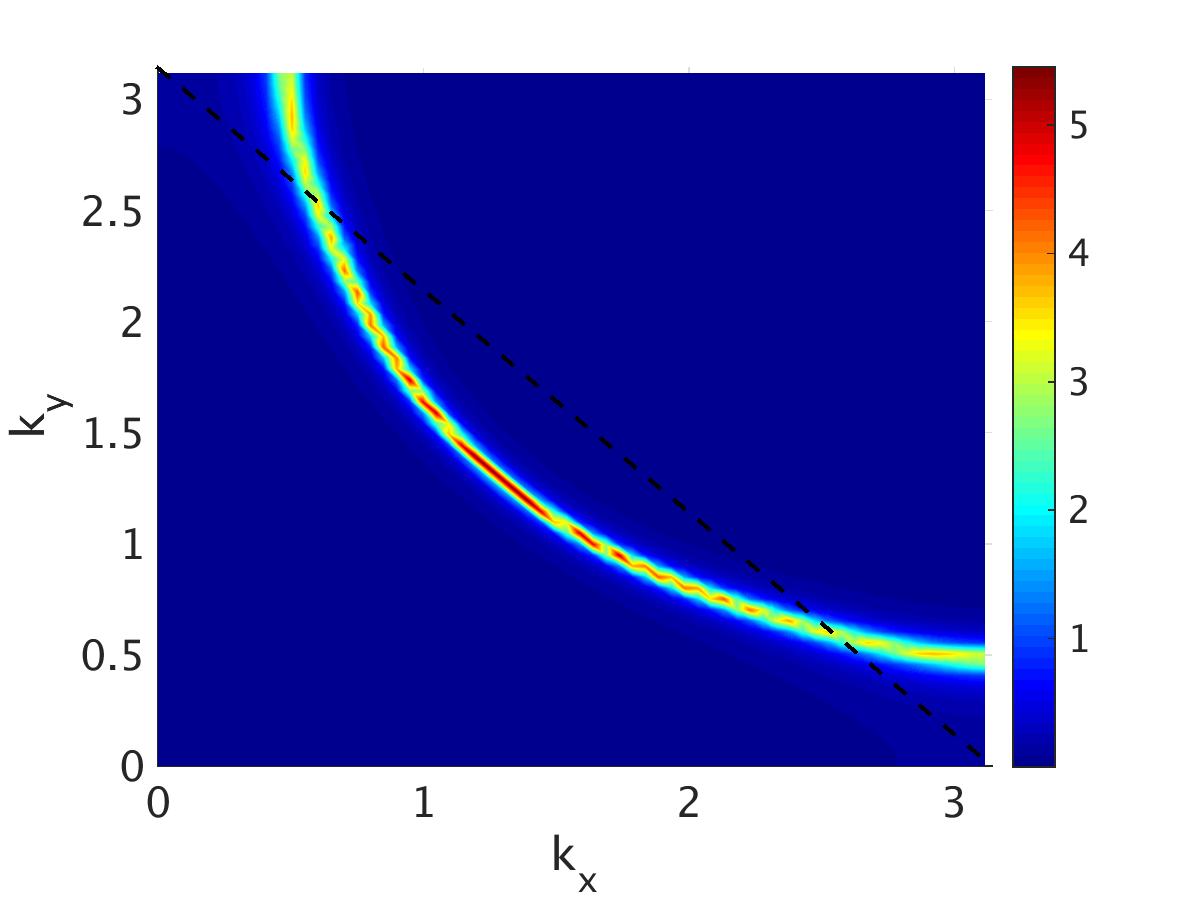}
\vspace*{-0.2cm}
%\vspace*{-0.3cm}
\caption{(Color online) Plot of interacting Fermi surface
  ($\rho_{\vk}(\omega=0)$) at strong coupling $g=5.0$ for $T=0.1,0.05,0.02$ (left to right.)}           
\label{fig:5}
\vspace*{-0.2cm}
\end{figure*}

\section{Results for dynamic model}
\subsection{Homogeneous situation}
As discussed in the last section, in the static limit self-energy corrections only lead to
a renormalization of the hoppings, but do not lead to any damping of
quasiparticles. For a dynamic interaction such as in Eq.~(\ref{eq:chidyn})
the effect is different.\cite{KS90a,KS90b,SPS98,SPS99}
% more references maybe
In fact, this one of the major aspects of
our work to include those effects on the single particle level. Scattering is 
particularly strong at the hot spots such that quasi-particle excitations can
become suppressed. Here, we would like to investigate the effect of this on the
charge order instabilities. The basic questions are: How much is the
diagonal order suppressed by the self-energy effects? Can the axial order
become favorable over the diagonal one?
In order to get some insights about the effects on the Fermi surface, we first
consider the homogeneous situation and solve the Eliashberg equations self-consistently
on the real frequency axis, Eq.~(\ref{eq:imsigmaw}). 
For clarity we first show results for weak interaction $g=1$. The other model
parameters are $t_1^0=1$, $t_2^0=-0.32$, $t_3^0=0.11$, $\mu_0=-1.1$, $\omega_{\rm
  sf}=0.5$, $a_{v_s}=1$, $\Gamma=0.5$, and $T=0.05$. The result for the
interacting Fermi surface, $\rho_{\vk}(\omega=0)$, for one part of the
Brillouin zone is shown in Fig.~\ref{fig:4} (left).

A broadening value $\eta=0.08$ was used to enhance clarity for the given $\vk$-space
resolution. Not surprisingly, 
the result resembles very much the non-interacting Fermi surface. On the right
in Fig.~\ref{fig:4} we also display $\rho_{\vk}(\omega)$ for
$\vk=(k_x,k_y)$ on the trajectory $(0,0)\to (\pi,\pi)\to (0,\pi)\to (0,0)$
($\Gamma\to M \to X \to \Gamma$), which gives insights on the damping and
renormalized band structure. We have added the non-interacting dispersion as a
dashed line. Here again the result is almost identical with the
non-interacting case.

We also show the results for a stronger interacting case $g=5$ in
Fig.~\ref{fig:5}, where $\mu=-1$ and we show the temperatures
$T=0.1,0.05,0.02$. Here, $\eta=0.05$ was chosen in the analytical continuation
for the broadening.

As we can see in Fig.~\ref{fig:5} the thermal part of the spin fluctuation spectrum
can give considerable effects on the electronic spectrum and give features
resemblant of Fermi arcs, which have been observed in experiments.\cite{Sea05} The
effect is more pronounced for larger temperature as can also be deduced
from Eq.~(\ref{eq:imsigmaw}). Hence, spectral weight is suppressed
close to the hot spots where the scattering with the spin-fluctuations is
particularly strong. Stronger arc-like features can be realized for larger
values of the correlation length $\xi$, i.e., smaller values of $\Gamma$. It
is worth noting that non-selfconsistent calculations also give stronger arc
features. However, self-consistent Eliashberg equations generically do not
produce real pseudogap features.\cite{SPS98,SPS99,CPS08}

\begin{figure}[!ht]
%\vspace*{-0.5cm}
\centering
\includegraphics[width=0.95\linewidth]{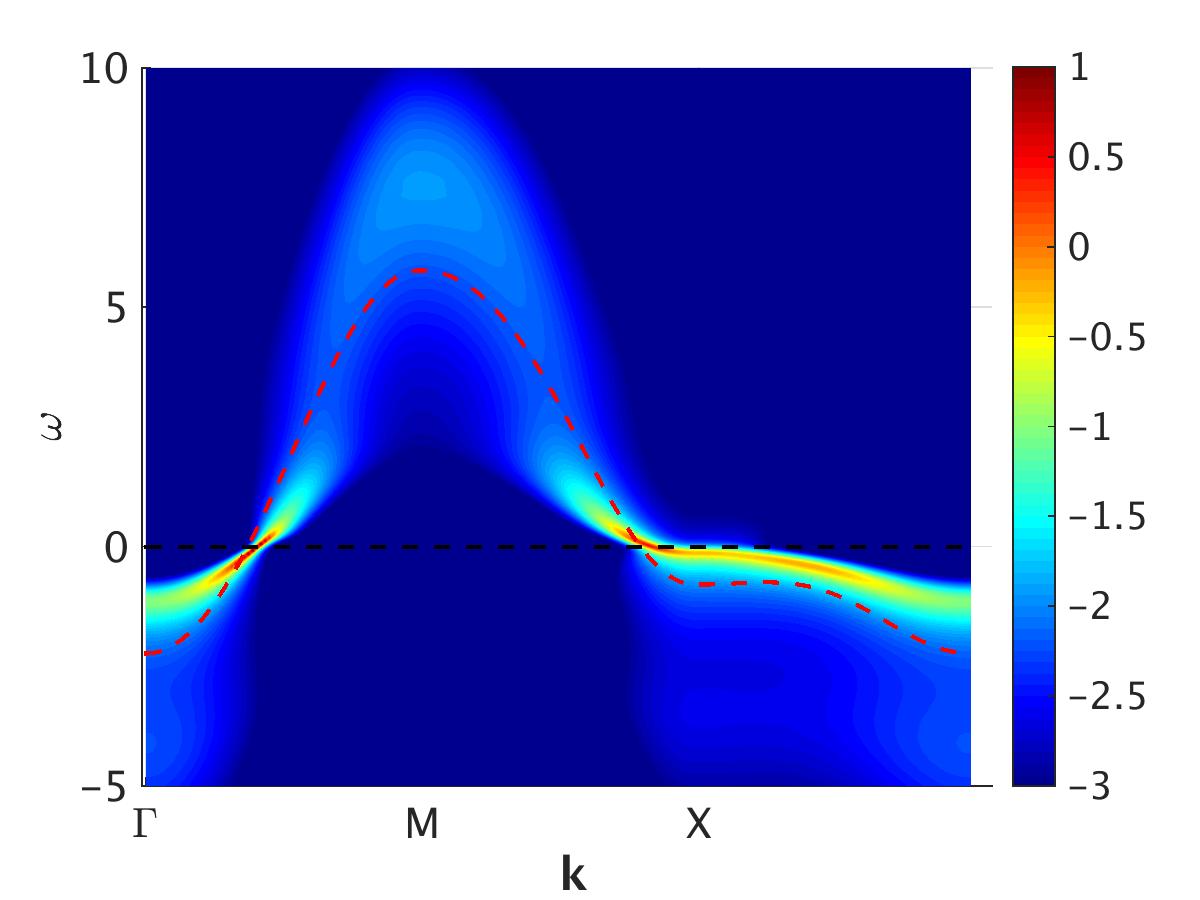}
\vspace*{-0.2cm}
%\vspace*{-0.3cm}
\caption{(Color online) Plot of renormalized band structure
  ($\log\rho_{\vk}(\omega)$ for clarity) at strong coupling $g=5.0$. Other
  model parameter as described in the text.}            
\label{fig:6}
\vspace*{-0.2cm}
\end{figure}

In Fig.~\ref{fig:6} we show for $T=0.05$ the renormalized band structure. 
We find effects of Fermi velocity renormalization and redistribution of
spectral weight. As we will discuss below this has consequences for the
instability analysis and realization of ordered  phases.

\begin{figure*}[!ht]
%\vspace*{-0.5cm}
\centering
\includegraphics[width=0.48\linewidth]{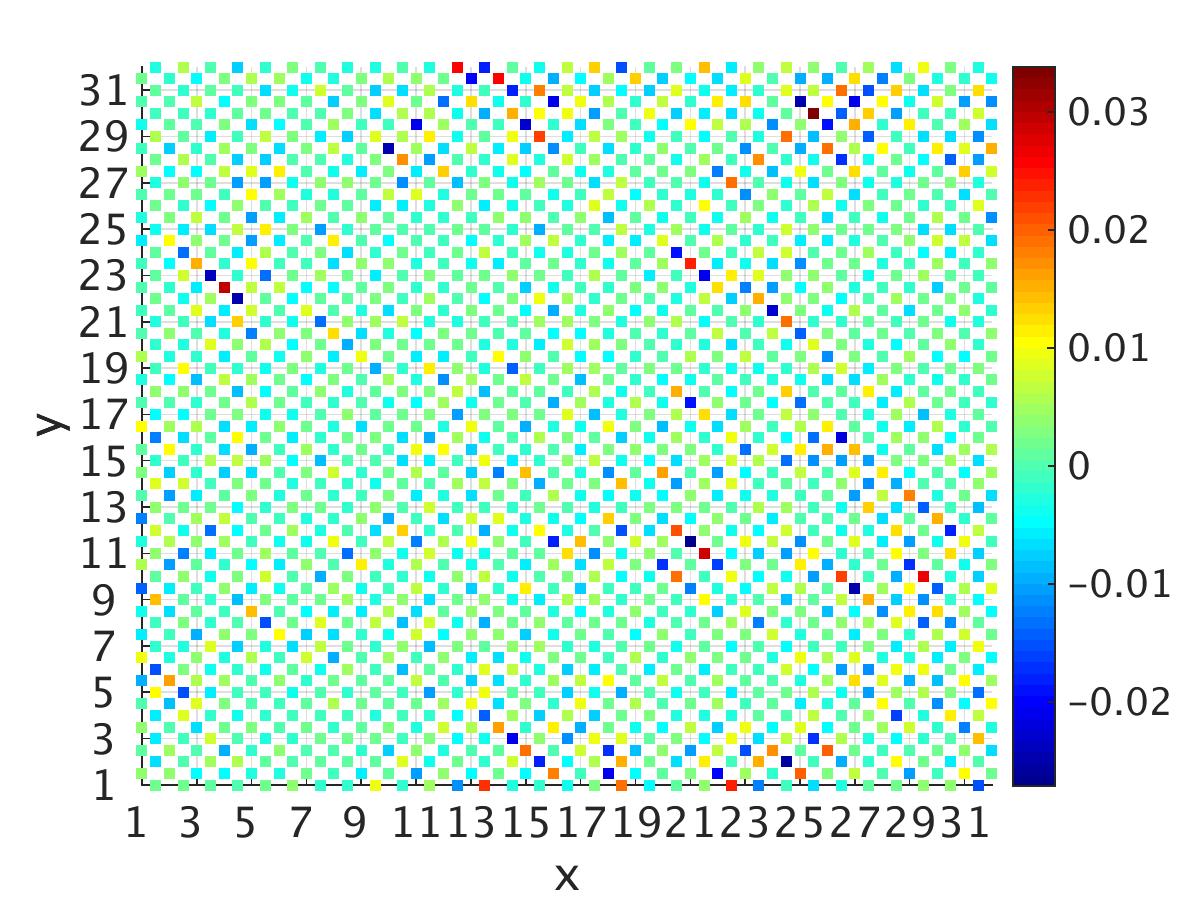}
\includegraphics[width=0.48\linewidth]{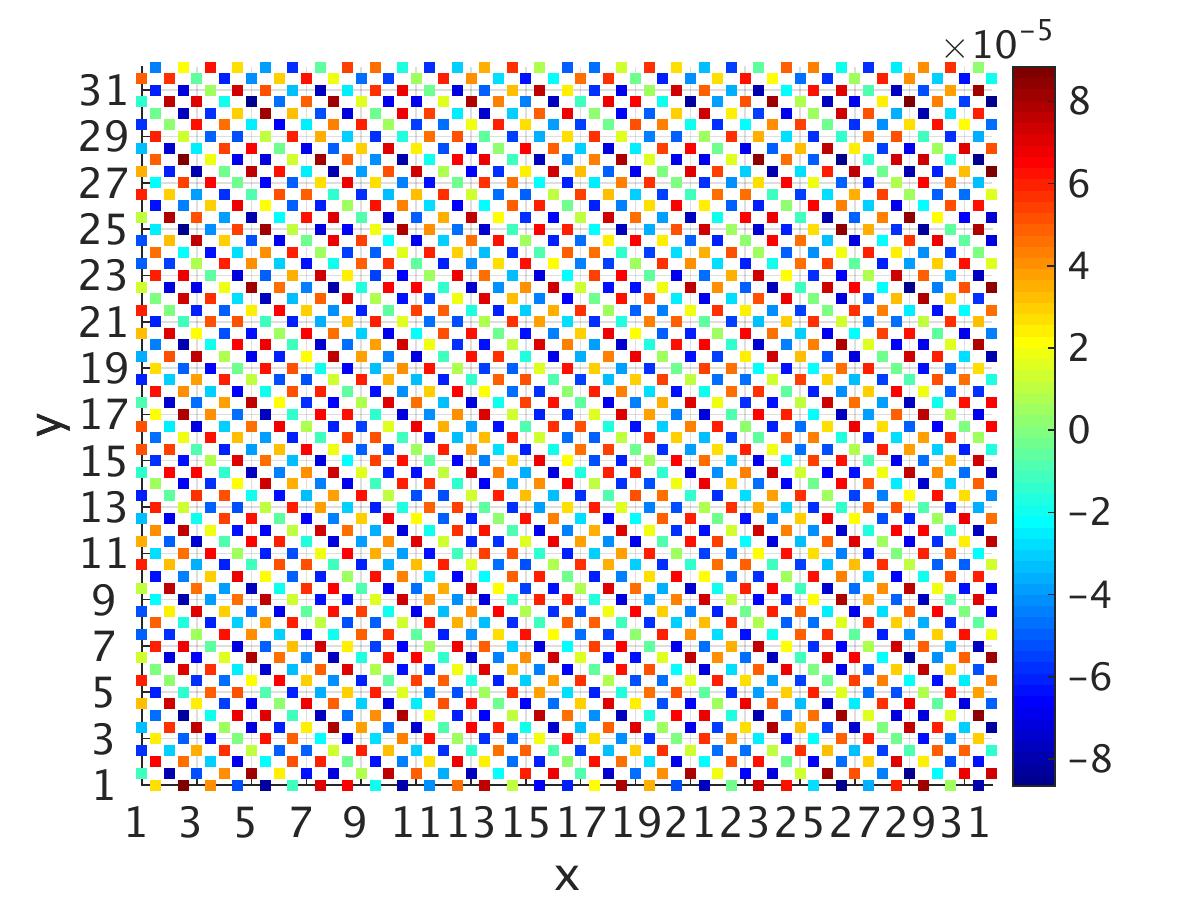}
\vspace*{-0.2cm}
%\vspace*{-0.3cm}
\caption{(Color online) Nearest neighbor bond values of $\Delta_{ij}^{\rm
    ih}$ on real space lattice initialized by a random field after 5
  iterations (left) and spontaneously ordered results after 45 
  iterations (right). Model parameter as described in the text.}            
\label{fig:7}
\vspace*{-0.2cm}
\end{figure*}

\subsection{Spontaneous order}
Within a similar procedure as described in Sec.~\ref{sec:static}, we have
done extensive calculations to check for charge order solution for the model
with the dynamic interaction. The calculations were done for a real space
lattice with $N_1=32$ and a grid of Matsubara frequency large enough to
capture relevant features. A random field $h_{ij}$ is used for initialization. In the
appendix, we describe some results which were obtained from calculations with field induced order.
For most calculations we used the hopping parameters $t_1^0=1$, $t_2^0=-0.32$, and
$t_3^0=0.11$. 
By varying $\mu_0$, we analyzed parameters with filling factors $n\sim 0.8-0.95$
and temperatures down to $T=0.02$ in units of $t_1^0$. The parameters of the
spin fluctuation spectrum where varied in a regime $\omega_{\rm sf}\sim
0.2-1$, and $\Gamma=1/2$ and $\Gamma=1/4$ was analyzed; we usually kept
$a_{v_s}=a_{\chi}=1$. We also scanned a range of coupling strengths
$g$. 

Generally, we found the appearance of the charge order to be strongly
suppressed in a scheme with dynamic spin fluctuations as compared to the static case. One reason
is the finite extent of the interaction in frequency space in contrast to the
static case. Another reason is
the single-particle renormalization effect of the self-energy. Charge order
enters via an inhomogeneous modulation of $\Sigma_{ij}(i\omega_n)$ in the
self-consistent calculation, which appears on top of generic variation of
$\Sigma_{ij}(i\omega_n)$ as function of $\vct r_i-\vct r_j$. This ordering
tendency should be enhanced with the coupling $g$. However, the self-energy
also has an effect to damp and renormalize single particle excitation, which
leads to the opposite effect of a suppression of the order.
This is a generic feature, which also appears for other Fermi surface
instabilities, such as superconductivity. As a consequence for many of the
probed parameters no charge ordered solution could be stabilized, even though
some ordering features appear in intermediate values of the iterations. For
instance, for $T=0.05$ we did not find spontaneously ordered solutions for
any of the parameters tested.

\begin{figure}[htbp]
%\vspace*{-0.5cm}
\centering
\includegraphics[width=0.48\textwidth]{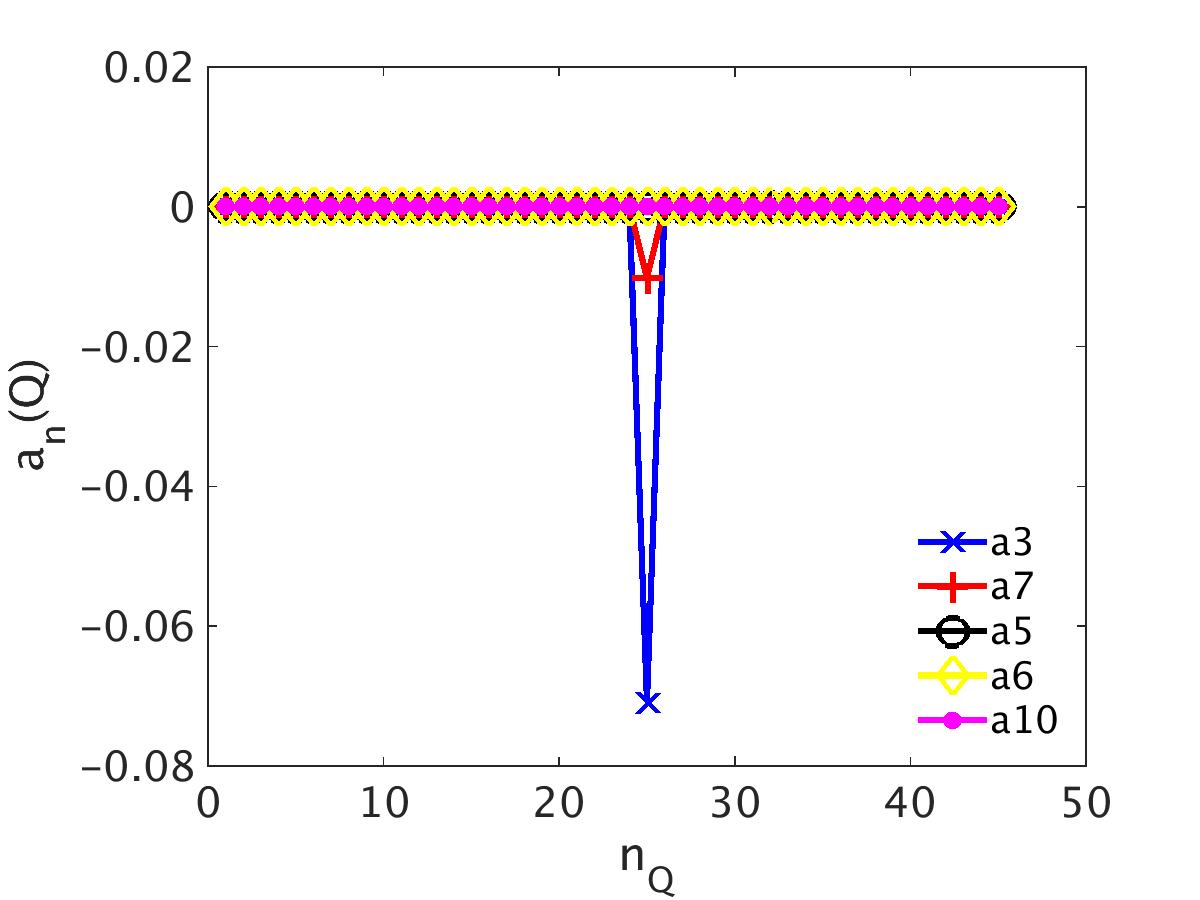}
\vspace*{-0.4cm}
%\vspace*{-0.3cm}
\caption{(Color online) Plot of the largest $a_n(\vct Q)$ for different basis functions
  ($n=1,\ldots, 13$). The values of $\vct Q$ are label by $n_{\vct Q}$, where
  $n_{\vct Q}=25$ corresponds to $\vct Q=Q_0(1 ,1)$, with $Q_0=3/8$ in units
  $\pi/a$. This value agrees well with the diagonal distance between hot
  spots.}           
\label{fig:8}
\vspace*{-0.3cm}
\end{figure}

In Fig.~\ref{fig:7}, we present the result of a calculation at $T=0.02$, where
spontaneous charge order appears in the self-consistent Eliashberg equations. The
parameters are $\omega_{\rm sf}=0.5$, $\Gamma=0.5$, with a coupling
$g=5$. Since $\chi(\vct r_i-\vct r_j,i\omega_m)$ is fairly small this coupling strength is
not as large as it might appear; for instance, the nearest neighbor self-energy
reaches maximal values $|\Real \Sigma(\vct r_i,\vct r_i+\hat x,i\omega_n)|\sim
0.2$ for these parameters. We show a result for $\mu_0=-1$, which correspond
to a filling of $n\simeq 0.92$ for the interacting system. The calculations
are again initialized by a random field and the spontaneous ordering pattern
visible after 45 iterations is seen on the right in Fig.~\ref{fig:7}. 
The corresponding Fourier analysis is displayed in Fig.~\ref{fig:8}.

As in the static case we find a dominant $d$-form factor, $n=3$ and a diagonal
wave vector with magnitude $\vct Q=Q_0(1, 1)$, with $Q_0=3/8$. There is also a
minor component for $n=7$, where $f_7(\vk)=\cos(2k_x) - \cos(2k_y)$. 
The wave vector matches the connection of the hot spots of the interacting
Fermi surface well (as seen in Fig.~\ref{fig:5}).
The diagonal order is dominant even though the spectral density is
substantially suppressed near the hot spots. 
For different filling factors, we also find diagonal order, whenever a hot spot is
present in the interacting Fermi surface, and the length of the wave vector
changes accordingly. 

In situations where there are no hot spots visible in the interacting Fermi
surface, for instance, when the filling is lowered such the Fermi surface
becomes closed around $(0,0)$, we found solutions with ordering tendencies
with wave vector along the axis. The dominant form factor is $n=3$, the $d$-form
factor. For filling $T=0.02$, $n\sim 0.8$ the wave vector $\vct Q=Q_0(1, 0)$, with
$Q_0=1/2$ can be found, which is very similar to the experimental wave length
$\lambda\sim 4$. The order is, however, quite small and not well 
established. It is possible that lower temperatures and additional ingredients are needed
to stabilize the order well.

\section{Discussion and Conclusions}

Based on a model of fermions on a square lattice interacting with dynamic
antiferromagnetic interactions, we have analyzed general charge order solutions. 
We used self-consistent Eliashberg equations to compute self-energies and spectral
functions. We showed that the spectral functions can be substantially modified
in the interacting theory. In particular, structures resembling Fermi arcs can
appear due to the strong scattering around the hot spots. However, no real
pseudogap features are generated by self-consistent Eliashberg equations.

The appearance of charge order is studied based on a real space version of the
Eliashberg equations, which are initialized by a random field. 
We only found spontaneously ordered solutions at low temperature, {\em e.g.\/},
$T=0.02 t_1$, and for relatively large interaction strengths. Generically, the
charge order posses a $d$-form factor and a diagonal wave vector related to the
hot spots of the interacting Fermi surface. This result is in line with a
number of previous studies,\cite{KKNUZ15,SL13,WC14,MN15pre,SS14,ABS14a,ABS14b} many
of which however did not include dynamic interactions and self-energy effects,
and studied instabilities from the normal 
state rather than ordered solutions. One of our main results is that 
the suppression of spectral weight at the hot spots by the dynamic model is not
large enough to alter the behavior of the charge order. For large dopings, when the interacting
Fermi surface has no hot spots, we found tendencies towards order with wave
vector along the axis. This finding is in line with studies suppressing the
hot spots, for instance, by pocket formation or including a phenomenological
pseudogap.\cite{BAK13,AKB15,TS15,ZM14pre,CS14b}

In relation to the observation of charge order in the cuprate superconductors,
we find a number of features consistent with the experiments. These include
the dominant $d$-form factor and the behavior of $\vct Q$ with doping. The
direction of the wave vector is, however, not correctly reproduced. This is
likely related to the fact that the self-consistent Eliashberg theory does not
include strong enough pseudogap features, and therefore still supports an
instability related to the hot spot. As such, the present approach does not provide a comprehensive
explanation for the experimental findings. Nevertheless, with suitable modifications
our real space Eliashberg approach might useful for future applications, as it
can easily be extended to include superconductivity, magnetic field effects,
impurity effects, coupling to phonons and spin order.

%%%%
%%%

\paragraph*{Acknowledgments -} We wish to thank A. Allais, D. Chowdhury, A.J. Millis, M. Norman,
M. Punk, P. Strack for fruitful discussions. JB would like to thank the
Deutsche Forschungsgemeinschaft for support through grant number BA 4371/1-1. 
This research was supported by the NSF under Grant DMR-1360789, the Templeton
foundation, and MURI grant W911NF-14-1-0003 from ARO. Research at Perimeter
Institute is supported by the Government of Canada through Industry Canada
and by the Province of Ontario through the Ministry of Research and
Innovation.

\begin{appendix}
\section*{Appendix}

\subsection{Basis functions}
Here we collect the basis functions which were used for the charge
order analysis. We focused on the 13 basis functions
$f_n(\vk)$ as shown in Table~\ref{tab:basisfunc}.
\begin{table}
  \centering
\begin{tabular}{c|c}
$n$ & $f_n(\vk)$         \\
\hline
1      & 1                         \\
2      & $\cos k_x + \cos k_y$      \\
3      & $\cos k_x - \cos k_y$      \\
4      & $\sin k_x + \sin k_y$      \\
5      & $\sin k_x - \sin k_y$      \\
6      & $\cos(2k_x) + \cos(2k_y)$  \\
7      & $\cos(2k_x) - \cos(2k_y)$  \\
8      & $\sin(2k_x) + \sin(2k_y)$  \\
9      & $\sin(2k_x) - \sin(2k_y)$  \\
10     & $2 \cos k_x \cos k_y$      \\
11     & $2 \sin k_x \sin k_y$      \\
12     & $2 \sin k_x \cos k_y$      \\
13     & $2 \cos k_x \sin k_y$      \\
\end{tabular}\hspace{20pt}
  \caption{Relevant basis functions}
  \label{tab:basisfunc}
\end{table}
The corresponding function of the real space representation $f_n(\vct r)$ can
be easily calculated analytically from Eq.~(\ref{eq:fnr}). They include form
factors up to next-nearest neighbors. Higher order functions could be in principle be considered but they
do not play a role in our analysis.  These functions $f_n(\vct r)$ are used to
compute the coefficients $a_n(\vct Q)$ in Eq.~(\ref{eq:anQ}). 

\subsection{Field induced order}
Here, we briefly summarize some results from a complementary analysis to Sec.~IV.
We studied charge order when a small finite field is present in the
self-consistent calculations. 
We focused on the situation with $d$-form factor and  assumed $\phi_3(\vct
Q)=\phi_3(-\vct Q)$. 
We chose $\phi_3(\vct Q)=0.001$ keeping the field finite in (\ref{eq:dysonre})
during the self-consistent calculation and scanned over various vectors $\vct
Q$. A response function characteristic of the susceptility to charge order can
be defined as $\chi(n,\vct Q)=\frac{\delta a_n(\vct Q)}{\delta \phi_n(\vct
  Q)}$.   

First of all we can consider the situation for $g=0$, such that
$\Sigma_{ij}=0$. Clearly there cannot be spontaneous order in this situation,
however, the susceptibility for different wave vectors can be analyzed. We find
that the field induced order as measured by $\chi^0(3,\vct Q)$ is strongest for
$\vct Q_d=(Q_0,Q_0)$, where $Q_0$ approximately connects the hot spots, and
there is also a local maximum along the $Q_x$ axis, for $\vct
Q_a=(Q_0,0)$. Within the real space calculations with $N_1=32$ one finds 
$\chi^0(3,\vct Q_a)/\chi^0(3,\vct Q_d)\simeq  0.78$. Hence the diagonal order
is favored. These results are clearly expected from the instability analysis
based on unrenormalized fermions\cite{SL13} where these quantities can be
computed from the corresponding fermionic bubbles and form factors.  

For finite $g$ the self-energy effects play an important role. As discussed
the ordering tendency is induced form an inhomogeneous contribution of
the real part of $\Sigma_{ij}(i\omega_n)$. 
In contrast to the static case $\Sigma_{ij}(i\omega_n)\to 0$  for large
$i\omega_n$ so this is only a contribution at small $\omega_n$. However, the
self-energy also produces other effects, such   as renormalization of the band
and suppression and shift of spectral weight as already discussed in the
section on the homogeneous calculations. 
These effects mostly lead to a reduction of the ordering
tendency. This is in fact common for weak coupling instabilities which are
strongest without self-energy corrections. 
Hence, the combination of these effects means that the field induced ordering
susceptility is not necessarily enhanced for finite $g$.
In our analysis for different parameters and interactions we found that
typically $\chi(3,\vct Q_d)$ with the diagonal wave-vector is largest and can
be enhanced over the non-interacting value. For the parameters studied
diagonal response was found to be larger than the $\chi(3,\vct Q_a)$, i.e.,
instabilities along the axial direction, consistent with what has been
discussed in the main text.

\end{appendix}

\bibliography{artikel}

\begin{thebibliography}{56}
\expandafter\ifx\csname natexlab\endcsname\relax\def\natexlab#1{#1}\fi
\expandafter\ifx\csname bibnamefont\endcsname\relax
  \def\bibnamefont#1{#1}\fi
\expandafter\ifx\csname bibfnamefont\endcsname\relax
  \def\bibfnamefont#1{#1}\fi
\expandafter\ifx\csname citenamefont\endcsname\relax
  \def\citenamefont#1{#1}\fi
\expandafter\ifx\csname url\endcsname\relax
  \def\url#1{\texttt{#1}}\fi
\expandafter\ifx\csname urlprefix\endcsname\relax\def\urlprefix{URL }\fi
\providecommand{\bibinfo}[2]{#2}
\providecommand{\eprint}[2][]{\url{#2}}

\bibitem[{\citenamefont{Fradkin and Kivelson}(2012)}]{FK12}
\bibinfo{author}{\bibfnamefont{E.}~\bibnamefont{Fradkin}} \bibnamefont{and}
  \bibinfo{author}{\bibfnamefont{S.~A.} \bibnamefont{Kivelson}},
  \bibinfo{journal}{Nature Physics} \textbf{\bibinfo{volume}{8}},
  \bibinfo{pages}{864} (\bibinfo{year}{2012}).

\bibitem[{\citenamefont{Keimer et~al.}(2015)\citenamefont{Keimer, Kivelson,
  Norman, Uchida, and Zaanen}}]{KKNUZ15}
\bibinfo{author}{\bibfnamefont{B.}~\bibnamefont{Keimer}},
  \bibinfo{author}{\bibfnamefont{S.}~\bibnamefont{Kivelson}},
  \bibinfo{author}{\bibfnamefont{M.}~\bibnamefont{Norman}},
  \bibinfo{author}{\bibfnamefont{S.}~\bibnamefont{Uchida}}, \bibnamefont{and}
  \bibinfo{author}{\bibfnamefont{J.}~\bibnamefont{Zaanen}},
  \bibinfo{journal}{Nature} \textbf{\bibinfo{volume}{518}},
  \bibinfo{pages}{179} (\bibinfo{year}{2015}).

\bibitem[{\citenamefont{Tranquada et~al.}(1995)\citenamefont{Tranquada,
  Sternlieb, Axe, Nakamura, and Uchida}}]{TSANU95}
\bibinfo{author}{\bibfnamefont{J.}~\bibnamefont{Tranquada}},
  \bibinfo{author}{\bibfnamefont{B.}~\bibnamefont{Sternlieb}},
  \bibinfo{author}{\bibfnamefont{J.}~\bibnamefont{Axe}},
  \bibinfo{author}{\bibfnamefont{Y.}~\bibnamefont{Nakamura}}, \bibnamefont{and}
  \bibinfo{author}{\bibfnamefont{S.}~\bibnamefont{Uchida}},
  \bibinfo{journal}{Nature} \textbf{\bibinfo{volume}{375}},
  \bibinfo{pages}{561} (\bibinfo{year}{1995}).

\bibitem[{\citenamefont{Emery et~al.}(1999)\citenamefont{Emery, Kivelson, and
  Tranquada}}]{EKT99}
\bibinfo{author}{\bibfnamefont{V.~J.} \bibnamefont{Emery}},
  \bibinfo{author}{\bibfnamefont{S.~A.} \bibnamefont{Kivelson}},
  \bibnamefont{and} \bibinfo{author}{\bibfnamefont{J.~M.}
  \bibnamefont{Tranquada}}, \bibinfo{journal}{Proc. Natl. Acad. Sci. (USA)}
  \textbf{\bibinfo{volume}{96}}, \bibinfo{pages}{8814} (\bibinfo{year}{1999}).

\bibitem[{\citenamefont{Kivelson et~al.}(2003)\citenamefont{Kivelson, Bindloss,
  Fradkin, Oganesyan, Tranquada, Kapitulnik, and Howald}}]{KBFOTKH03}
\bibinfo{author}{\bibfnamefont{S.~A.} \bibnamefont{Kivelson}},
  \bibinfo{author}{\bibfnamefont{I.~P.} \bibnamefont{Bindloss}},
  \bibinfo{author}{\bibfnamefont{E.}~\bibnamefont{Fradkin}},
  \bibinfo{author}{\bibfnamefont{V.}~\bibnamefont{Oganesyan}},
  \bibinfo{author}{\bibfnamefont{J.~M.} \bibnamefont{Tranquada}},
  \bibinfo{author}{\bibfnamefont{A.}~\bibnamefont{Kapitulnik}},
  \bibnamefont{and} \bibinfo{author}{\bibfnamefont{C.}~\bibnamefont{Howald}},
  \bibinfo{journal}{Rev. Mod. Phys.} \textbf{\bibinfo{volume}{75}},
  \bibinfo{pages}{1201} (\bibinfo{year}{2003}).

\bibitem[{\citenamefont{Hoffman et~al.}(2002)\citenamefont{Hoffman, Hudson,
  Lang, Madhavan, Eisaki, Uchida, and Davis}}]{HHLMEUD02}
\bibinfo{author}{\bibfnamefont{J.}~\bibnamefont{Hoffman}},
  \bibinfo{author}{\bibfnamefont{E.}~\bibnamefont{Hudson}},
  \bibinfo{author}{\bibfnamefont{K.}~\bibnamefont{Lang}},
  \bibinfo{author}{\bibfnamefont{V.}~\bibnamefont{Madhavan}},
  \bibinfo{author}{\bibfnamefont{H.}~\bibnamefont{Eisaki}},
  \bibinfo{author}{\bibfnamefont{S.}~\bibnamefont{Uchida}}, \bibnamefont{and}
  \bibinfo{author}{\bibfnamefont{J.}~\bibnamefont{Davis}},
  \bibinfo{journal}{Science} \textbf{\bibinfo{volume}{295}},
  \bibinfo{pages}{466} (\bibinfo{year}{2002}).

\bibitem[{\citenamefont{Vershinin et~al.}(2004)\citenamefont{Vershinin, Misra,
  Ono, Abe, Ando, and Yazdani}}]{VMOAAY04}
\bibinfo{author}{\bibfnamefont{M.}~\bibnamefont{Vershinin}},
  \bibinfo{author}{\bibfnamefont{S.}~\bibnamefont{Misra}},
  \bibinfo{author}{\bibfnamefont{S.}~\bibnamefont{Ono}},
  \bibinfo{author}{\bibfnamefont{Y.}~\bibnamefont{Abe}},
  \bibinfo{author}{\bibfnamefont{Y.}~\bibnamefont{Ando}}, \bibnamefont{and}
  \bibinfo{author}{\bibfnamefont{A.}~\bibnamefont{Yazdani}},
  \bibinfo{journal}{Science} \textbf{\bibinfo{volume}{303}},
  \bibinfo{pages}{1995} (\bibinfo{year}{2004}).

\bibitem[{\citenamefont{Kohsaka et~al.}(2007)\citenamefont{Kohsaka, Taylor,
  Fujita, Schmidt, Lupien, Hanaguri, Azuma, Takano, Eisaki, Takagi
  et~al.}}]{Kea07}
\bibinfo{author}{\bibfnamefont{Y.}~\bibnamefont{Kohsaka}},
  \bibinfo{author}{\bibfnamefont{C.}~\bibnamefont{Taylor}},
  \bibinfo{author}{\bibfnamefont{K.}~\bibnamefont{Fujita}},
  \bibinfo{author}{\bibfnamefont{A.}~\bibnamefont{Schmidt}},
  \bibinfo{author}{\bibfnamefont{C.}~\bibnamefont{Lupien}},
  \bibinfo{author}{\bibfnamefont{T.}~\bibnamefont{Hanaguri}},
  \bibinfo{author}{\bibfnamefont{M.}~\bibnamefont{Azuma}},
  \bibinfo{author}{\bibfnamefont{M.}~\bibnamefont{Takano}},
  \bibinfo{author}{\bibfnamefont{H.}~\bibnamefont{Eisaki}},
  \bibinfo{author}{\bibfnamefont{H.}~\bibnamefont{Takagi}},
  \bibnamefont{et~al.}, \bibinfo{journal}{Science}
  \textbf{\bibinfo{volume}{315}}, \bibinfo{pages}{1380} (\bibinfo{year}{2007}).

\bibitem[{\citenamefont{Lawler et~al.}(2010)\citenamefont{Lawler, Fujita, Lee,
  Schmidt, Kohsaka, Kim, Eisaki, Uchida, Davis, Sethna et~al.}}]{Lea10}
\bibinfo{author}{\bibfnamefont{M.}~\bibnamefont{Lawler}},
  \bibinfo{author}{\bibfnamefont{K.}~\bibnamefont{Fujita}},
  \bibinfo{author}{\bibfnamefont{J.}~\bibnamefont{Lee}},
  \bibinfo{author}{\bibfnamefont{A.}~\bibnamefont{Schmidt}},
  \bibinfo{author}{\bibfnamefont{Y.}~\bibnamefont{Kohsaka}},
  \bibinfo{author}{\bibfnamefont{C.~K.} \bibnamefont{Kim}},
  \bibinfo{author}{\bibfnamefont{H.}~\bibnamefont{Eisaki}},
  \bibinfo{author}{\bibfnamefont{S.}~\bibnamefont{Uchida}},
  \bibinfo{author}{\bibfnamefont{J.}~\bibnamefont{Davis}},
  \bibinfo{author}{\bibfnamefont{J.}~\bibnamefont{Sethna}},
  \bibnamefont{et~al.}, \bibinfo{journal}{Nature}
  \textbf{\bibinfo{volume}{466}}, \bibinfo{pages}{347} (\bibinfo{year}{2010}).

\bibitem[{\citenamefont{Ghiringhelli et~al.}(2012)\citenamefont{Ghiringhelli,
  Le~Tacon, Minola, Blanco-Canosa, Mazzoli, Brookes, De~Luca, Frano, Hawthorn,
  He et~al.}}]{Gea12}
\bibinfo{author}{\bibfnamefont{G.}~\bibnamefont{Ghiringhelli}},
  \bibinfo{author}{\bibfnamefont{M.}~\bibnamefont{Le~Tacon}},
  \bibinfo{author}{\bibfnamefont{M.}~\bibnamefont{Minola}},
  \bibinfo{author}{\bibfnamefont{S.}~\bibnamefont{Blanco-Canosa}},
  \bibinfo{author}{\bibfnamefont{C.}~\bibnamefont{Mazzoli}},
  \bibinfo{author}{\bibfnamefont{N.}~\bibnamefont{Brookes}},
  \bibinfo{author}{\bibfnamefont{G.}~\bibnamefont{De~Luca}},
  \bibinfo{author}{\bibfnamefont{A.}~\bibnamefont{Frano}},
  \bibinfo{author}{\bibfnamefont{D.}~\bibnamefont{Hawthorn}},
  \bibinfo{author}{\bibfnamefont{F.}~\bibnamefont{He}}, \bibnamefont{et~al.},
  \bibinfo{journal}{Science} \textbf{\bibinfo{volume}{337}},
  \bibinfo{pages}{821} (\bibinfo{year}{2012}).

\bibitem[{\citenamefont{Chang et~al.}(2012)\citenamefont{Chang, Blackburn,
  Holmes, Christensen, Larsen, Mesot, Liang, Bonn, Hardy, Watenphul
  et~al.}}]{Cea12}
\bibinfo{author}{\bibfnamefont{J.}~\bibnamefont{Chang}},
  \bibinfo{author}{\bibfnamefont{E.}~\bibnamefont{Blackburn}},
  \bibinfo{author}{\bibfnamefont{A.}~\bibnamefont{Holmes}},
  \bibinfo{author}{\bibfnamefont{N.}~\bibnamefont{Christensen}},
  \bibinfo{author}{\bibfnamefont{J.}~\bibnamefont{Larsen}},
  \bibinfo{author}{\bibfnamefont{J.}~\bibnamefont{Mesot}},
  \bibinfo{author}{\bibfnamefont{R.}~\bibnamefont{Liang}},
  \bibinfo{author}{\bibfnamefont{D.}~\bibnamefont{Bonn}},
  \bibinfo{author}{\bibfnamefont{W.}~\bibnamefont{Hardy}},
  \bibinfo{author}{\bibfnamefont{A.}~\bibnamefont{Watenphul}},
  \bibnamefont{et~al.}, \bibinfo{journal}{Nature Physics}
  \textbf{\bibinfo{volume}{8}}, \bibinfo{pages}{871} (\bibinfo{year}{2012}).

\bibitem[{\citenamefont{Achkar et~al.}(2012)\citenamefont{Achkar, Sutarto, Mao,
  He, Frano, Blanco-Canosa, Le~Tacon, Ghiringhelli, Braicovich, Minola
  et~al.}}]{Aea12}
\bibinfo{author}{\bibfnamefont{A.~J.} \bibnamefont{Achkar}},
  \bibinfo{author}{\bibfnamefont{R.}~\bibnamefont{Sutarto}},
  \bibinfo{author}{\bibfnamefont{X.}~\bibnamefont{Mao}},
  \bibinfo{author}{\bibfnamefont{F.}~\bibnamefont{He}},
  \bibinfo{author}{\bibfnamefont{A.}~\bibnamefont{Frano}},
  \bibinfo{author}{\bibfnamefont{S.}~\bibnamefont{Blanco-Canosa}},
  \bibinfo{author}{\bibfnamefont{M.}~\bibnamefont{Le~Tacon}},
  \bibinfo{author}{\bibfnamefont{G.}~\bibnamefont{Ghiringhelli}},
  \bibinfo{author}{\bibfnamefont{L.}~\bibnamefont{Braicovich}},
  \bibinfo{author}{\bibfnamefont{M.}~\bibnamefont{Minola}},
  \bibnamefont{et~al.}, \bibinfo{journal}{Phys. Rev. Lett.}
  \textbf{\bibinfo{volume}{109}}, \bibinfo{pages}{167001}
  (\bibinfo{year}{2012}).

\bibitem[{\citenamefont{Wu et~al.}(2011)\citenamefont{Wu, Mayaffre, Kr{\"a}mer,
  Horvati{\'c}, Berthier, Hardy, Liang, Bonn, and Julien}}]{Wea11}
\bibinfo{author}{\bibfnamefont{T.}~\bibnamefont{Wu}},
  \bibinfo{author}{\bibfnamefont{H.}~\bibnamefont{Mayaffre}},
  \bibinfo{author}{\bibfnamefont{S.}~\bibnamefont{Kr{\"a}mer}},
  \bibinfo{author}{\bibfnamefont{M.}~\bibnamefont{Horvati{\'c}}},
  \bibinfo{author}{\bibfnamefont{C.}~\bibnamefont{Berthier}},
  \bibinfo{author}{\bibfnamefont{W.}~\bibnamefont{Hardy}},
  \bibinfo{author}{\bibfnamefont{R.}~\bibnamefont{Liang}},
  \bibinfo{author}{\bibfnamefont{D.}~\bibnamefont{Bonn}}, \bibnamefont{and}
  \bibinfo{author}{\bibfnamefont{M.-H.} \bibnamefont{Julien}},
  \bibinfo{journal}{Nature} \textbf{\bibinfo{volume}{477}},
  \bibinfo{pages}{191} (\bibinfo{year}{2011}).

\bibitem[{\citenamefont{LeBoeuf et~al.}(2013)\citenamefont{LeBoeuf, Kr{\"a}mer,
  Hardy, Liang, Bonn, and Proust}}]{LKHKBP13}
\bibinfo{author}{\bibfnamefont{D.}~\bibnamefont{LeBoeuf}},
  \bibinfo{author}{\bibfnamefont{S.}~\bibnamefont{Kr{\"a}mer}},
  \bibinfo{author}{\bibfnamefont{W.}~\bibnamefont{Hardy}},
  \bibinfo{author}{\bibfnamefont{R.}~\bibnamefont{Liang}},
  \bibinfo{author}{\bibfnamefont{D.}~\bibnamefont{Bonn}}, \bibnamefont{and}
  \bibinfo{author}{\bibfnamefont{C.}~\bibnamefont{Proust}},
  \bibinfo{journal}{Nature Physics} \textbf{\bibinfo{volume}{9}},
  \bibinfo{pages}{79} (\bibinfo{year}{2013}).

\bibitem[{\citenamefont{Doiron-Leyraud
  et~al.}(2007)\citenamefont{Doiron-Leyraud, Proust, LeBoeuf, Levallois,
  Bonnemaison, Liang, Bonn, Hardy, and Taillefer}}]{Dea07}
\bibinfo{author}{\bibfnamefont{N.}~\bibnamefont{Doiron-Leyraud}},
  \bibinfo{author}{\bibfnamefont{C.}~\bibnamefont{Proust}},
  \bibinfo{author}{\bibfnamefont{D.}~\bibnamefont{LeBoeuf}},
  \bibinfo{author}{\bibfnamefont{J.}~\bibnamefont{Levallois}},
  \bibinfo{author}{\bibfnamefont{J.-B.} \bibnamefont{Bonnemaison}},
  \bibinfo{author}{\bibfnamefont{R.}~\bibnamefont{Liang}},
  \bibinfo{author}{\bibfnamefont{D.}~\bibnamefont{Bonn}},
  \bibinfo{author}{\bibfnamefont{W.}~\bibnamefont{Hardy}}, \bibnamefont{and}
  \bibinfo{author}{\bibfnamefont{L.}~\bibnamefont{Taillefer}},
  \bibinfo{journal}{Nature} \textbf{\bibinfo{volume}{447}},
  \bibinfo{pages}{565} (\bibinfo{year}{2007}).

\bibitem[{\citenamefont{Sebastian et~al.}(2012)\citenamefont{Sebastian,
  Harrison, and Lonzarich}}]{SHL12}
\bibinfo{author}{\bibfnamefont{S.~E.} \bibnamefont{Sebastian}},
  \bibinfo{author}{\bibfnamefont{N.}~\bibnamefont{Harrison}}, \bibnamefont{and}
  \bibinfo{author}{\bibfnamefont{G.~G.} \bibnamefont{Lonzarich}},
  \bibinfo{journal}{Rep. Prog. Phys.} \textbf{\bibinfo{volume}{75}},
  \bibinfo{pages}{102501} (\bibinfo{year}{2012}).

\bibitem[{\citenamefont{Allais et~al.}(2014{\natexlab{a}})\citenamefont{Allais,
  Chowdhury, and Sachdev}}]{ACS14}
\bibinfo{author}{\bibfnamefont{A.}~\bibnamefont{Allais}},
  \bibinfo{author}{\bibfnamefont{D.}~\bibnamefont{Chowdhury}},
  \bibnamefont{and} \bibinfo{author}{\bibfnamefont{S.}~\bibnamefont{Sachdev}},
  \bibinfo{journal}{Nature communications} \textbf{\bibinfo{volume}{5}}
  (\bibinfo{year}{2014}{\natexlab{a}}).

\bibitem[{\citenamefont{Comin et~al.}(2014)\citenamefont{Comin, Frano, Yee,
  Yoshida, Eisaki, Schierle, Weschke, Sutarto, He, Soumyanarayanan
  et~al.}}]{Cea14a}
\bibinfo{author}{\bibfnamefont{R.}~\bibnamefont{Comin}},
  \bibinfo{author}{\bibfnamefont{A.}~\bibnamefont{Frano}},
  \bibinfo{author}{\bibfnamefont{M.}~\bibnamefont{Yee}},
  \bibinfo{author}{\bibfnamefont{Y.}~\bibnamefont{Yoshida}},
  \bibinfo{author}{\bibfnamefont{H.}~\bibnamefont{Eisaki}},
  \bibinfo{author}{\bibfnamefont{E.}~\bibnamefont{Schierle}},
  \bibinfo{author}{\bibfnamefont{E.}~\bibnamefont{Weschke}},
  \bibinfo{author}{\bibfnamefont{R.}~\bibnamefont{Sutarto}},
  \bibinfo{author}{\bibfnamefont{F.}~\bibnamefont{He}},
  \bibinfo{author}{\bibfnamefont{A.}~\bibnamefont{Soumyanarayanan}},
  \bibnamefont{et~al.}, \bibinfo{journal}{Science}
  \textbf{\bibinfo{volume}{343}}, \bibinfo{pages}{390} (\bibinfo{year}{2014}).

\bibitem[{\citenamefont{da~Silva~Neto et~al.}(2014)\citenamefont{da~Silva~Neto,
  Aynajian, Frano, Comin, Schierle, Weschke, Gyenis, Wen, Schneeloch, Xu
  et~al.}}]{Dea14a}
\bibinfo{author}{\bibfnamefont{E.~H.} \bibnamefont{da~Silva~Neto}},
  \bibinfo{author}{\bibfnamefont{P.}~\bibnamefont{Aynajian}},
  \bibinfo{author}{\bibfnamefont{A.}~\bibnamefont{Frano}},
  \bibinfo{author}{\bibfnamefont{R.}~\bibnamefont{Comin}},
  \bibinfo{author}{\bibfnamefont{E.}~\bibnamefont{Schierle}},
  \bibinfo{author}{\bibfnamefont{E.}~\bibnamefont{Weschke}},
  \bibinfo{author}{\bibfnamefont{A.}~\bibnamefont{Gyenis}},
  \bibinfo{author}{\bibfnamefont{J.}~\bibnamefont{Wen}},
  \bibinfo{author}{\bibfnamefont{J.}~\bibnamefont{Schneeloch}},
  \bibinfo{author}{\bibfnamefont{Z.}~\bibnamefont{Xu}}, \bibnamefont{et~al.},
  \bibinfo{journal}{Science} \textbf{\bibinfo{volume}{343}},
  \bibinfo{pages}{393} (\bibinfo{year}{2014}).

\bibitem[{\citenamefont{H\"ucker et~al.}(2011)\citenamefont{H\"ucker,
  v.~Zimmermann, Gu, Xu, Wen, Xu, Kang, Zheludev, and Tranquada}}]{Hea11}
\bibinfo{author}{\bibfnamefont{M.}~\bibnamefont{H\"ucker}},
  \bibinfo{author}{\bibfnamefont{M.}~\bibnamefont{v.~Zimmermann}},
  \bibinfo{author}{\bibfnamefont{G.~D.} \bibnamefont{Gu}},
  \bibinfo{author}{\bibfnamefont{Z.~J.} \bibnamefont{Xu}},
  \bibinfo{author}{\bibfnamefont{J.~S.} \bibnamefont{Wen}},
  \bibinfo{author}{\bibfnamefont{G.}~\bibnamefont{Xu}},
  \bibinfo{author}{\bibfnamefont{H.~J.} \bibnamefont{Kang}},
  \bibinfo{author}{\bibfnamefont{A.}~\bibnamefont{Zheludev}}, \bibnamefont{and}
  \bibinfo{author}{\bibfnamefont{J.~M.} \bibnamefont{Tranquada}},
  \bibinfo{journal}{Phys. Rev. B} \textbf{\bibinfo{volume}{83}},
  \bibinfo{pages}{104506} (\bibinfo{year}{2011}).

\bibitem[{\citenamefont{Tranquada}(2013)}]{Tra13}
\bibinfo{author}{\bibfnamefont{J.~M.} \bibnamefont{Tranquada}}, in
  \emph{\bibinfo{booktitle}{American Institute of Physics Conference Series}}
  (\bibinfo{year}{2013}), vol. \bibinfo{volume}{1550}, pp.
  \bibinfo{pages}{114--187}.

\bibitem[{\citenamefont{Fujita et~al.}(2014)\citenamefont{Fujita, Hamidian,
  Edkins, Kim, Kohsaka, Azuma, Takano, Takagi, Eisaki, Uchida et~al.}}]{Fea14}
\bibinfo{author}{\bibfnamefont{K.}~\bibnamefont{Fujita}},
  \bibinfo{author}{\bibfnamefont{M.~H.} \bibnamefont{Hamidian}},
  \bibinfo{author}{\bibfnamefont{S.~D.} \bibnamefont{Edkins}},
  \bibinfo{author}{\bibfnamefont{C.~K.} \bibnamefont{Kim}},
  \bibinfo{author}{\bibfnamefont{Y.}~\bibnamefont{Kohsaka}},
  \bibinfo{author}{\bibfnamefont{M.}~\bibnamefont{Azuma}},
  \bibinfo{author}{\bibfnamefont{M.}~\bibnamefont{Takano}},
  \bibinfo{author}{\bibfnamefont{H.}~\bibnamefont{Takagi}},
  \bibinfo{author}{\bibfnamefont{H.}~\bibnamefont{Eisaki}},
  \bibinfo{author}{\bibfnamefont{S.-i.} \bibnamefont{Uchida}},
  \bibnamefont{et~al.}, \bibinfo{journal}{Proc. Nat. Acad. Sci.}
  \textbf{\bibinfo{volume}{111}}, \bibinfo{pages}{E3026}
  (\bibinfo{year}{2014}).

\bibitem[{\citenamefont{Comin et~al.}(2015)\citenamefont{Comin, Sutarto, He,
  da~Silva~Neto, Chauviere, Frano, Liang, Hardy, Bonn, Yoshida et~al.}}]{Cea15}
\bibinfo{author}{\bibfnamefont{R.}~\bibnamefont{Comin}},
  \bibinfo{author}{\bibfnamefont{R.}~\bibnamefont{Sutarto}},
  \bibinfo{author}{\bibfnamefont{F.}~\bibnamefont{He}},
  \bibinfo{author}{\bibfnamefont{E.~H.} \bibnamefont{da~Silva~Neto}},
  \bibinfo{author}{\bibfnamefont{L.}~\bibnamefont{Chauviere}},
  \bibinfo{author}{\bibfnamefont{A.}~\bibnamefont{Frano}},
  \bibinfo{author}{\bibfnamefont{R.}~\bibnamefont{Liang}},
  \bibinfo{author}{\bibfnamefont{W.~N.} \bibnamefont{Hardy}},
  \bibinfo{author}{\bibfnamefont{D.~A.} \bibnamefont{Bonn}},
  \bibinfo{author}{\bibfnamefont{Y.}~\bibnamefont{Yoshida}},
  \bibnamefont{et~al.}, \bibinfo{journal}{Nat. Mater.}
  \textbf{\bibinfo{volume}{advance online publication}} (\bibinfo{year}{2015}),
  \urlprefix\url{http://dx.doi.org/10.1038/nmat4295}.

\bibitem[{\citenamefont{Hamidian et~al.}(2015)\citenamefont{Hamidian, Edkins,
  Kim, Davis, Mackenzie, Eisaki, Uchida, Lawler, Kim, Sachdev
  et~al.}}]{Hea15pre}
\bibinfo{author}{\bibfnamefont{M.~H.} \bibnamefont{Hamidian}},
  \bibinfo{author}{\bibfnamefont{S.~D.} \bibnamefont{Edkins}},
  \bibinfo{author}{\bibfnamefont{C.~K.} \bibnamefont{Kim}},
  \bibinfo{author}{\bibfnamefont{J.~C.} \bibnamefont{Davis}},
  \bibinfo{author}{\bibfnamefont{A.~P.} \bibnamefont{Mackenzie}},
  \bibinfo{author}{\bibfnamefont{H.}~\bibnamefont{Eisaki}},
  \bibinfo{author}{\bibfnamefont{S.}~\bibnamefont{Uchida}},
  \bibinfo{author}{\bibfnamefont{M.~J.} \bibnamefont{Lawler}},
  \bibinfo{author}{\bibfnamefont{E.-A.} \bibnamefont{Kim}},
  \bibinfo{author}{\bibfnamefont{S.}~\bibnamefont{Sachdev}},
  \bibnamefont{et~al.}, \bibinfo{journal}{to appear}  (\bibinfo{year}{2015}).

\bibitem[{\citenamefont{Holder and Metzner}(2012)}]{HM12a}
\bibinfo{author}{\bibfnamefont{T.}~\bibnamefont{Holder}} \bibnamefont{and}
  \bibinfo{author}{\bibfnamefont{W.}~\bibnamefont{Metzner}},
  \bibinfo{journal}{Phys. Rev. B} \textbf{\bibinfo{volume}{85}},
  \bibinfo{pages}{165130} (\bibinfo{year}{2012}).

\bibitem[{\citenamefont{Husemann and Metzner}(2012)}]{HM12b}
\bibinfo{author}{\bibfnamefont{C.}~\bibnamefont{Husemann}} \bibnamefont{and}
  \bibinfo{author}{\bibfnamefont{W.}~\bibnamefont{Metzner}},
  \bibinfo{journal}{Phys. Rev. B} \textbf{\bibinfo{volume}{86}},
  \bibinfo{pages}{085113} (\bibinfo{year}{2012}).

\bibitem[{\citenamefont{Bejas et~al.}(2012)\citenamefont{Bejas, Greco, and
  Yamase}}]{BGY12}
\bibinfo{author}{\bibfnamefont{M.}~\bibnamefont{Bejas}},
  \bibinfo{author}{\bibfnamefont{A.}~\bibnamefont{Greco}}, \bibnamefont{and}
  \bibinfo{author}{\bibfnamefont{H.}~\bibnamefont{Yamase}},
  \bibinfo{journal}{Phys. Rev. B} \textbf{\bibinfo{volume}{86}},
  \bibinfo{pages}{224509} (\bibinfo{year}{2012}).

\bibitem[{\citenamefont{Sachdev and La~Placa}(2013)}]{SL13}
\bibinfo{author}{\bibfnamefont{S.}~\bibnamefont{Sachdev}} \bibnamefont{and}
  \bibinfo{author}{\bibfnamefont{R.}~\bibnamefont{La~Placa}},
  \bibinfo{journal}{Phys. Rev. Lett.} \textbf{\bibinfo{volume}{111}},
  \bibinfo{pages}{027202} (\bibinfo{year}{2013}).

\bibitem[{\citenamefont{Wang and Chubukov}(2014)}]{WC14}
\bibinfo{author}{\bibfnamefont{Y.}~\bibnamefont{Wang}} \bibnamefont{and}
  \bibinfo{author}{\bibfnamefont{A.}~\bibnamefont{Chubukov}},
  \bibinfo{journal}{Phys. Rev. B} \textbf{\bibinfo{volume}{90}},
  \bibinfo{pages}{035149} (\bibinfo{year}{2014}).

\bibitem[{\citenamefont{Maier and Scalapino}(2014)}]{MS14}
\bibinfo{author}{\bibfnamefont{T.~A.} \bibnamefont{Maier}} \bibnamefont{and}
  \bibinfo{author}{\bibfnamefont{D.~J.} \bibnamefont{Scalapino}},
  \bibinfo{journal}{Phys. Rev. B} \textbf{\bibinfo{volume}{90}},
  \bibinfo{pages}{174510} (\bibinfo{year}{2014}).

\bibitem[{\citenamefont{Tsvelik and Chubukov}(2014)}]{TC14}
\bibinfo{author}{\bibfnamefont{A.~M.} \bibnamefont{Tsvelik}} \bibnamefont{and}
  \bibinfo{author}{\bibfnamefont{A.~V.} \bibnamefont{Chubukov}},
  \bibinfo{journal}{Phys. Rev. B} \textbf{\bibinfo{volume}{89}},
  \bibinfo{pages}{184515} (\bibinfo{year}{2014}).

\bibitem[{\citenamefont{Lee}(2014)}]{Lee14}
\bibinfo{author}{\bibfnamefont{P.~A.} \bibnamefont{Lee}},
  \bibinfo{journal}{Phys. Rev. X} \textbf{\bibinfo{volume}{4}},
  \bibinfo{pages}{031017} (\bibinfo{year}{2014}).

\bibitem[{\citenamefont{de~Carvalho and Freire}(2014)}]{DF14}
\bibinfo{author}{\bibfnamefont{V.~S.} \bibnamefont{de~Carvalho}}
  \bibnamefont{and} \bibinfo{author}{\bibfnamefont{H.}~\bibnamefont{Freire}},
  \bibinfo{journal}{Annals of Physics} \textbf{\bibinfo{volume}{348}},
  \bibinfo{pages}{32 } (\bibinfo{year}{2014}), ISSN \bibinfo{issn}{0003-4916}.

\bibitem[{\citenamefont{{Mishra} and {Norman}}(2015)}]{MN15pre}
\bibinfo{author}{\bibfnamefont{V.}~\bibnamefont{{Mishra}}} \bibnamefont{and}
  \bibinfo{author}{\bibfnamefont{M.~R.} \bibnamefont{{Norman}}},
  \bibinfo{journal}{ArXiv e-prints}  (\bibinfo{year}{2015}),
  \eprint{1502.02782}.

\bibitem[{\citenamefont{Wang and Chubukov}(2015)}]{WC15}
\bibinfo{author}{\bibfnamefont{Y.}~\bibnamefont{Wang}} \bibnamefont{and}
  \bibinfo{author}{\bibfnamefont{A.}~\bibnamefont{Chubukov}},
  \bibinfo{journal}{Phys. Rev. B} \textbf{\bibinfo{volume}{91}},
  \bibinfo{pages}{195113} (\bibinfo{year}{2015}).

\bibitem[{\citenamefont{Metlitski and Sachdev}(2010)}]{MS10a}
\bibinfo{author}{\bibfnamefont{M.~A.} \bibnamefont{Metlitski}}
  \bibnamefont{and} \bibinfo{author}{\bibfnamefont{S.}~\bibnamefont{Sachdev}},
  \bibinfo{journal}{Phys. Rev. B} \textbf{\bibinfo{volume}{82}},
  \bibinfo{pages}{075128} (\bibinfo{year}{2010}).

\bibitem[{\citenamefont{Sau and Sachdev}(2014)}]{SS14}
\bibinfo{author}{\bibfnamefont{J.~D.} \bibnamefont{Sau}} \bibnamefont{and}
  \bibinfo{author}{\bibfnamefont{S.}~\bibnamefont{Sachdev}},
  \bibinfo{journal}{Phys. Rev. B} \textbf{\bibinfo{volume}{89}},
  \bibinfo{pages}{075129} (\bibinfo{year}{2014}).

\bibitem[{\citenamefont{Allais et~al.}(2014{\natexlab{b}})\citenamefont{Allais,
  Bauer, and Sachdev}}]{ABS14a}
\bibinfo{author}{\bibfnamefont{A.}~\bibnamefont{Allais}},
  \bibinfo{author}{\bibfnamefont{J.}~\bibnamefont{Bauer}}, \bibnamefont{and}
  \bibinfo{author}{\bibfnamefont{S.}~\bibnamefont{Sachdev}},
  \bibinfo{journal}{Indian Journal of Physics} pp. \bibinfo{pages}{1--9}
  (\bibinfo{year}{2014}{\natexlab{b}}), ISSN \bibinfo{issn}{0973-1458}.

\bibitem[{\citenamefont{Allais et~al.}(2014{\natexlab{c}})\citenamefont{Allais,
  Bauer, and Sachdev}}]{ABS14b}
\bibinfo{author}{\bibfnamefont{A.}~\bibnamefont{Allais}},
  \bibinfo{author}{\bibfnamefont{J.}~\bibnamefont{Bauer}}, \bibnamefont{and}
  \bibinfo{author}{\bibfnamefont{S.}~\bibnamefont{Sachdev}},
  \bibinfo{journal}{Phys. Rev. B} \textbf{\bibinfo{volume}{90}},
  \bibinfo{pages}{155114} (\bibinfo{year}{2014}{\natexlab{c}}).

\bibitem[{\citenamefont{Bulut et~al.}(2013)\citenamefont{Bulut, Atkinson, and
  Kampf}}]{BAK13}
\bibinfo{author}{\bibfnamefont{S.}~\bibnamefont{Bulut}},
  \bibinfo{author}{\bibfnamefont{W.~A.} \bibnamefont{Atkinson}},
  \bibnamefont{and} \bibinfo{author}{\bibfnamefont{A.~P.} \bibnamefont{Kampf}},
  \bibinfo{journal}{Phys. Rev. B} \textbf{\bibinfo{volume}{88}},
  \bibinfo{pages}{155132} (\bibinfo{year}{2013}).

\bibitem[{\citenamefont{Atkinson et~al.}(2015)\citenamefont{Atkinson, Kampf,
  and Bulut}}]{AKB15}
\bibinfo{author}{\bibfnamefont{W.}~\bibnamefont{Atkinson}},
  \bibinfo{author}{\bibfnamefont{A.}~\bibnamefont{Kampf}}, \bibnamefont{and}
  \bibinfo{author}{\bibfnamefont{S.}~\bibnamefont{Bulut}},
  \bibinfo{journal}{New Journal of Physics} \textbf{\bibinfo{volume}{17}},
  \bibinfo{pages}{013025} (\bibinfo{year}{2015}).

\bibitem[{\citenamefont{Thomson and Sachdev}(2015)}]{TS15}
\bibinfo{author}{\bibfnamefont{A.}~\bibnamefont{Thomson}} \bibnamefont{and}
  \bibinfo{author}{\bibfnamefont{S.}~\bibnamefont{Sachdev}},
  \bibinfo{journal}{Phys. Rev. B} \textbf{\bibinfo{volume}{91}},
  \bibinfo{pages}{115142} (\bibinfo{year}{2015}).

\bibitem[{\citenamefont{{Zhang} and {Mei}}(2014)}]{ZM14pre}
\bibinfo{author}{\bibfnamefont{L.}~\bibnamefont{{Zhang}}} \bibnamefont{and}
  \bibinfo{author}{\bibfnamefont{J.-W.} \bibnamefont{{Mei}}},
  \bibinfo{journal}{ArXiv e-prints}  (\bibinfo{year}{2014}),
  \eprint{1408.6592}.

\bibitem[{\citenamefont{Chowdhury and Sachdev}(2014{\natexlab{a}})}]{CS14a}
\bibinfo{author}{\bibfnamefont{D.}~\bibnamefont{Chowdhury}} \bibnamefont{and}
  \bibinfo{author}{\bibfnamefont{S.}~\bibnamefont{Sachdev}},
  \bibinfo{journal}{Phys. Rev. B} \textbf{\bibinfo{volume}{90}},
  \bibinfo{pages}{134516} (\bibinfo{year}{2014}{\natexlab{a}}).

\bibitem[{\citenamefont{P\'epin et~al.}(2014)\citenamefont{P\'epin,
  de~Carvalho, Kloss, and Montiel}}]{PCKM14}
\bibinfo{author}{\bibfnamefont{C.}~\bibnamefont{P\'epin}},
  \bibinfo{author}{\bibfnamefont{V.~S.} \bibnamefont{de~Carvalho}},
  \bibinfo{author}{\bibfnamefont{T.}~\bibnamefont{Kloss}}, \bibnamefont{and}
  \bibinfo{author}{\bibfnamefont{X.}~\bibnamefont{Montiel}},
  \bibinfo{journal}{Phys. Rev. B} \textbf{\bibinfo{volume}{90}},
  \bibinfo{pages}{195207} (\bibinfo{year}{2014}).

\bibitem[{\citenamefont{Wang et~al.}(2015)\citenamefont{Wang, Agterberg, and
  Chubukov}}]{WAC15}
\bibinfo{author}{\bibfnamefont{Y.}~\bibnamefont{Wang}},
  \bibinfo{author}{\bibfnamefont{D.~F.} \bibnamefont{Agterberg}},
  \bibnamefont{and} \bibinfo{author}{\bibfnamefont{A.}~\bibnamefont{Chubukov}},
  \bibinfo{journal}{Phys. Rev. B} \textbf{\bibinfo{volume}{91}},
  \bibinfo{pages}{115103} (\bibinfo{year}{2015}).

\bibitem[{\citenamefont{Punk}(2015)}]{Pun15}
\bibinfo{author}{\bibfnamefont{M.}~\bibnamefont{Punk}}, \bibinfo{journal}{Phys.
  Rev. B} \textbf{\bibinfo{volume}{91}}, \bibinfo{pages}{115131}
  (\bibinfo{year}{2015}).

\bibitem[{\citenamefont{Chowdhury and Sachdev}(2014{\natexlab{b}})}]{CS14b}
\bibinfo{author}{\bibfnamefont{D.}~\bibnamefont{Chowdhury}} \bibnamefont{and}
  \bibinfo{author}{\bibfnamefont{S.}~\bibnamefont{Sachdev}},
  \bibinfo{journal}{Phys. Rev. B} \textbf{\bibinfo{volume}{90}},
  \bibinfo{pages}{245136} (\bibinfo{year}{2014}{\natexlab{b}}).

\bibitem[{\citenamefont{Abanov et~al.}(2003)\citenamefont{Abanov, Chubukov, and
  Schmalian}}]{ACS03}
\bibinfo{author}{\bibfnamefont{A.}~\bibnamefont{Abanov}},
  \bibinfo{author}{\bibfnamefont{A.~V.} \bibnamefont{Chubukov}},
  \bibnamefont{and}
  \bibinfo{author}{\bibfnamefont{J.}~\bibnamefont{Schmalian}},
  \bibinfo{journal}{Advances in Physics} \textbf{\bibinfo{volume}{52}},
  \bibinfo{pages}{119} (\bibinfo{year}{2003}).

\bibitem[{\citenamefont{Chubukov et~al.}(2008)\citenamefont{Chubukov, Pines,
  and Schmalian}}]{CPS08}
\bibinfo{author}{\bibfnamefont{A.}~\bibnamefont{Chubukov}},
  \bibinfo{author}{\bibfnamefont{D.}~\bibnamefont{Pines}}, \bibnamefont{and}
  \bibinfo{author}{\bibfnamefont{J.}~\bibnamefont{Schmalian}}, in
  \emph{\bibinfo{booktitle}{Superconductivity (Vol 2)}}, edited by
  \bibinfo{editor}{\bibfnamefont{K.}~\bibnamefont{Bennemann}} \bibnamefont{and}
  \bibinfo{editor}{\bibfnamefont{J.}~\bibnamefont{Ketterson}}
  (\bibinfo{publisher}{Springer}, \bibinfo{address}{Berlin},
  \bibinfo{year}{2008}).

\bibitem[{\citenamefont{Millis et~al.}(1990)\citenamefont{Millis, Monien, and
  Pines}}]{MMP90}
\bibinfo{author}{\bibfnamefont{A.}~\bibnamefont{Millis}},
  \bibinfo{author}{\bibfnamefont{H.}~\bibnamefont{Monien}}, \bibnamefont{and}
  \bibinfo{author}{\bibfnamefont{D.}~\bibnamefont{Pines}},
  \bibinfo{journal}{Phys. Rev. B} \textbf{\bibinfo{volume}{42}},
  \bibinfo{pages}{167} (\bibinfo{year}{1990}).

\bibitem[{\citenamefont{Kampf and Schrieffer}(1990{\natexlab{a}})}]{KS90a}
\bibinfo{author}{\bibfnamefont{A.}~\bibnamefont{Kampf}} \bibnamefont{and}
  \bibinfo{author}{\bibfnamefont{J.~R.} \bibnamefont{Schrieffer}},
  \bibinfo{journal}{Phys. Rev. B} \textbf{\bibinfo{volume}{41}},
  \bibinfo{pages}{6399} (\bibinfo{year}{1990}{\natexlab{a}}).

\bibitem[{\citenamefont{Kampf and Schrieffer}(1990{\natexlab{b}})}]{KS90b}
\bibinfo{author}{\bibfnamefont{A.~P.} \bibnamefont{Kampf}} \bibnamefont{and}
  \bibinfo{author}{\bibfnamefont{J.~R.} \bibnamefont{Schrieffer}},
  \bibinfo{journal}{Phys. Rev. B} \textbf{\bibinfo{volume}{42}},
  \bibinfo{pages}{7967} (\bibinfo{year}{1990}{\natexlab{b}}).

\bibitem[{\citenamefont{Schmalian et~al.}(1998)\citenamefont{Schmalian, Pines,
  and Stojkovi\ifmmode~\acute{c}\else \'{c}\fi{}}}]{SPS98}
\bibinfo{author}{\bibfnamefont{J.}~\bibnamefont{Schmalian}},
  \bibinfo{author}{\bibfnamefont{D.}~\bibnamefont{Pines}}, \bibnamefont{and}
  \bibinfo{author}{\bibfnamefont{B.}~\bibnamefont{Stojkovi\ifmmode~\acute{c}\else
  \'{c}\fi{}}}, \bibinfo{journal}{Phys. Rev. Lett.}
  \textbf{\bibinfo{volume}{80}}, \bibinfo{pages}{3839} (\bibinfo{year}{1998}).

\bibitem[{\citenamefont{Schmalian et~al.}(1999)\citenamefont{Schmalian, Pines,
  and Stojkovi\ifmmode~\acute{c}\else \'{c}\fi{}}}]{SPS99}
\bibinfo{author}{\bibfnamefont{J.}~\bibnamefont{Schmalian}},
  \bibinfo{author}{\bibfnamefont{D.}~\bibnamefont{Pines}}, \bibnamefont{and}
  \bibinfo{author}{\bibfnamefont{B.}~\bibnamefont{Stojkovi\ifmmode~\acute{c}\else
  \'{c}\fi{}}}, \bibinfo{journal}{Phys. Rev. B} \textbf{\bibinfo{volume}{60}},
  \bibinfo{pages}{667} (\bibinfo{year}{1999}).

\bibitem[{\citenamefont{Shen et~al.}(2005)\citenamefont{Shen, Ronning, Lu,
  Baumberger, Ingle, Lee, Meevasana, Kohsaka, Azuma, Takano et~al.}}]{Sea05}
\bibinfo{author}{\bibfnamefont{K.~M.} \bibnamefont{Shen}},
  \bibinfo{author}{\bibfnamefont{F.}~\bibnamefont{Ronning}},
  \bibinfo{author}{\bibfnamefont{D.}~\bibnamefont{Lu}},
  \bibinfo{author}{\bibfnamefont{F.}~\bibnamefont{Baumberger}},
  \bibinfo{author}{\bibfnamefont{N.}~\bibnamefont{Ingle}},
  \bibinfo{author}{\bibfnamefont{W.}~\bibnamefont{Lee}},
  \bibinfo{author}{\bibfnamefont{W.}~\bibnamefont{Meevasana}},
  \bibinfo{author}{\bibfnamefont{Y.}~\bibnamefont{Kohsaka}},
  \bibinfo{author}{\bibfnamefont{M.}~\bibnamefont{Azuma}},
  \bibinfo{author}{\bibfnamefont{M.}~\bibnamefont{Takano}},
  \bibnamefont{et~al.}, \bibinfo{journal}{Science}
  \textbf{\bibinfo{volume}{307}}, \bibinfo{pages}{901} (\bibinfo{year}{2005}).

\end{thebibliography}

\end{document}